\documentclass[11pt,a4paper]{article}
\usepackage[utf8]{inputenc}
\usepackage{graphicx}
\usepackage{amssymb} \usepackage{amsmath} \usepackage{graphicx}
\usepackage{epsfig,latexsym,color,xcolor}
\usepackage{ulem}
\usepackage{amstext}    
\usepackage{array} 
\usepackage{slashed}
\usepackage[height=10in ,hmargin={1.6cm,0.8in}]{geometry}
\usepackage{braket}
\usepackage{appendix}
\usepackage{authblk}

\DeclareUnicodeCharacter{0229}{\c{e}}

\newcommand\numberthis{\stepcounter{equation}\tag{\theequation}}

\makeatletter
\newcommand*{\rom}[1]{\expandafter\@slowromancap\romannumeral #1@}
\makeatother

\renewcommand{\theequation}{\thesection.\arabic{equation}}
\numberwithin{equation}{section}

\begin{document}
\title{Accelerating Strangelets via Penrose process in non-BPS fuzzballs}
\date{\today}
\author[1]{Massimo Bianchi}
\author[1]{Marco Casolino}
\author[1]{Gabriele Rizzo}
\affil[1]{\it Dipartimento di Fisica, Universit\`a di Roma Tor Vergata, I.N.F.N. Sezione di Roma Tor Vergata, Via della Ricerca Scientifica, 1 - 00133 Roma, ITALY}

\maketitle

\begin{abstract}
Ultra High Energy Cosmic Rays may include {\it strangelets}, a form of Strange Quark Matter, among their components. We briefly review their properties and discuss how they can be accelerated via Penrose process taking place in singular rotating Kerr black holes or in their smooth, horizonless counterparts in string theory, according to the {\it fuzzball} proposal. We focus on non-BPS solutions of the JMaRT kind and compute the efficiency of Penrose process that turns out not to be bounded unlike for Kerr BHs.
\end{abstract}

\tableofcontents 

\newpage

\section{Introduction and motivations}
Cosmic rays (CR) and in particular Ultra High Energy CR (UHECR) tend to play an important role in any progress of high-energy physics, from the identification of new elementary particles in the past to the recent confirmation of rare phenomena such as neutrino oscillations. 

Although the flux is tremendously suppressed at very high energies \cite{Abbasi:2007sv, Abraham:2008ru, Abraham:2010mj} and practically ends at the ZeV scale set by the GZK cutoff~\cite{Greisen:1966jv, Zatsepin:1966jv}, the question remains as for how UHECR can be accelerated up to such high energies. 

Among the components of UHECR one can include the so-called ``strangelets'', a form of Strange Quark Matter (SQM) \cite{witten, DeRujula1984, Madsen2004} that can be present in the dense core of a Neutron Star (NS) or in Quark Stars (QS) \cite{ivanenko}, where the temperature and density may be significantly higher than on the crust. When the mass of a NS exceeds the Chandrasekhar-Oppenheimer-Volkoff bound (around a few solar masses), it becomes unstable wrt gravitational collapse and produces a Black Hole (BH). In turn, BHs may provide both tidal tearing of captured astrophysical objects, including NS and QS, and a powerful acceleration mechanism of UHECR, including strangelets, aka Penrose process \cite{Penrose:1971uk}. This can take place in rotating (Kerr) BHs surrounded by an `ergo-region' where a time-like Killing vector becomes space-like. Thanks to Penrose mechanism Kerr/rotating  BHs can be used as cosmic slings to accelerate UHECR and reach the GZK cutoff scale. 

BHs are the epitome of quantum gravity (QG), which is still poorly understood with quantum field-theory means. Luckily there is a leading contender: string theory. Based on the idea that point-like particles be replaced by one-dimensional objects, string theory can accommodate gravity (mediated by closed strings) together with gauge interactions (mediated by open strings) in an entirely consistent framework. General Relativity or higher dimensional extensions thereof, coupled to gauge fields and fermions, governs the dynamics at very low energies, compared to the string mass scale.

The realization that string theory in addition to fundamental strings admits stable, extended solitons with $p$ spatial dimensions, called $p$-branes, allows to represent BHs as bound states of strings and branes and to quantitatively address and partly solve some long-standing issues in the physics of BHs \cite{BH-branes}, including BH production in high-energy collisions \cite{BHprod}. In particular there are classes of charged BPS\footnote{A BPS state (after Bogomolny, Prasad and Sommerfield) is an extremal state that saturates a (supersymmetric) bound between mass $M$, charges $Q$ and angular momentum $J$.} BHs for which one can precisely count the micro-states responsible for the macroscopic entropy, which according to Beckenstein and Hawking is proportional to the area of the event horizon \cite{StromVafaetc}. In the emerging `fuzz-ball' proposal  \cite{FuzzMathur}, BHs are described as ensembles of smooth horizonless geometries with the same asymptotic behavior at infinity, that have a non-trivial structure at the putative horizon. By replacing BHs with fuzzballs, dense, tangled balls of strings some of the subtle paradoxes can be avoided or clarified since they were generated by accepting the very presence of the singularity and of the horizon that must be only an approximation valid in the classical limit.

The aim of the present paper is to use non-BPS fuzz-balls with an ergo-region as cosmic slings for the acceleration of UHECR. Due to a no-go theorem, that prevents the existence of non-trivial smooth horizonless solutions in $D=4$ \cite{NoGoFuzz4}, we have to rely on fuzz-balls in higher dimensions \cite{Fuzz5&6}. In particular, $D=5$ and $D=6$ will be our starting point and represent a toy model for the physically interesting case. We will mostly use JMaRT (after Jejjalla, Madden, Ross and Titchener \cite{JMaRT}) solitonic solution that is smooth and horizonless, yet with an ergo-region. As we will see, the asymptotic geometry, though free from pathological Closed Time-like Curves (CTC's), is over-rotating and cannot be as such strictly identified with the fuzz-ball of a BH, not even in $D=5$ \cite{MoreOnJMaRT}. In fact JMaRT displays an instability that suggests that this kind of charged non-BPS solutions should decay into BPS ones with the same charges and lower angular momenta such as to satisfy the bound for BHs. The instability of JMaRT has been addressed by various groups, including \cite{JMaRTinstab}\footnote{For a similar analysis in the BPS context see \cite{BPSinstab}.}. To the best of our knowledge however the role of the Penrose process in JMaRT or in similar smooth, horizonless solitonic geometries have not been addressed previously\footnote{We thank G.~Bossard and D.~Turton for confirming this.}. 

The plan of the paper is as follows.

In Section 2 we will briefly review the notion of Strange Quark Matter and strangelets, discuss how Penrose mechanism can have a role in accelerating UHECRs including strangelets and SQM and sketch the alternative acceleration mechanisms proposed so far.

In Section 3 we will recall the  JMaRT solution and its properties. We set the stage for the analysis of the Penrose process with the study of geodesics motion in JMaRT geometry. Thanks to the large amount of isometry it proves convenient to work in the Hamiltonian formulation. We focus on the geodesics in the $\theta = 0$ hyperplane and compute the efficiency of the Penrose process for massive spin-less particles that in-fall in a counter-rotating way. We discuss the results in comparison with the analogous process for Kerr BHs, reviewed in an Appendix. 

Section 4 contains our conclusions and an outlook for future investigation in this subject. In particular we will comment on upper limits that MINI-EUSO can set on the flux of strangelets and on the Penrose mechanism for their acceleration derived in Section 3 for non-BPS fuzz-balls and reviewed in the Appendix.

In the Appendix, for the sake of convenience and for comparison with our analysis for non-BPS fuzz-balls, we briefly review the rotating BH solution, originally found by Kerr, and its properties and discuss the Penrose process for both massive particles decaying into a massless pair (photons). 

\section{Strangelets and their acceleration via Penrose process}

Neutron stars (NS) are probably the most compact and dense form of `ordinary' matter in the Universe \cite{Weinberg, Dyson}. When thermo-nuclear reactions have exhausted their fuel, protons tend to recombine with electrons and form neutrons via emission of neutrinos. Matter becomes so dense that in a radius of a few km one can package masses of the order of the Sun's. In fact modelling the system as a perfect fluid with spherical symmetry Tolman, Oppenheimer and Volkoff (TOV) wrote down relativistic equations that allow determining quantitative bounds on the masses of these compact stars \cite{Weinberg, Dyson}. Assuming very low temperatures so much so that NS be supported only by the pressure of the degenerate Fermi gas of neutrons, one can set an upper limit on the mass of NS around the mass of the Sun. {\it Mutatis mutandis} i.e. replacing electrons with neutrons and using a `reasonable' equation of state (EoS), relating pressure and energy density, the result is strikingly similar to Chandrasekar bound on the mass of white dwarves \cite{Chandra, Weinberg, Shutz}. More recently, stimulated by the observation of candidate NS's violating the bound through direct GW detection \cite{GWfromNS}, more elaborate equations of states have been proposed that amount to slice NS radially with different layers satisfying different EoS's that are glued at the interface \cite{Weinberg, Dyson}. In particular it has been suggested that Strange Quark Matter (SQM) \cite{witten}, composed of up $u$, down $d$ and strange $s$ quarks, may play a role in internal layers, whereby temperature can raise and deconfinement can take place thus giving rise to a quark-gluon plasma. In fact the concept of a Quark Star (QS) has been put forward \cite{ivanenko}.

\subsection{Strange Quark Matter and strangelets}

The existence of SQM as a different state of hadronic matter other than ordinary nuclear matter was proposed for the first time in 1984 \cite{witten}. 
SQM would be composed by roughly an equal number of $u$, $d$ and $s$ quarks, with the presence of a third quark lowering  the  nucleon Fermi level with respect to a system with only two quark flavours \cite{madsen2}. In this case SQM may constitute the true ground state of hadronic matter and be stable. Quarks would  be lumped together and not separated in nucleons, resulting in quark matter being  much denser than ordinary matter.  

SQM could have been produced in the Big Bang \cite{madsen}, be present in the core of neutron stars or in ``Strange Quark Stars'' (SQS) \cite{ivanenko} and be a candidate for  baryonic dark matter \cite{2014AIPC.1604..121T}. Portions of SQM could be ejected as a consequence of collisions of these stars in binary systems \cite{drago}. Such collisions can inject a small fraction of this matter (also called strangelets) in the galactic radiation where it could be identifiable with cosmic ray detectors or mass spectrometers.
Various experiments have tried to produce or search for SQM in various environments, on the ground, on balloons and in satellites, both of active and passive nature.   A  review on strangelet search and models can be found in \cite{Finch_2006, scientificreports2017}.

SQM should be neutral (uncharged), if an exactly equal number of $u$, $d$, and $s$ quarks is dynamically favoured, however the neutrality condition may be approximate, allowing strangelets to have a small residual electrical charge. 
In the light mass range, these objects could be identified as having an anomalous $A/Z>>2$ ratio. 
A search with the PAMELA  space-borne magnetic spectrometer has yielded upper limits $\simeq 2\cdot 10^3 p/(m^2\ sr\ yr)$ in the mass range $2<A<10^5$ \cite{pamela}.

In \cite{DeRujula1984} it has been suggested that heavier objects could interact with the atmosphere through an adiabatic compression mechanism similar to that of meteors. The higher density of SQM would result in longer and more uniform  tracks than in case of meteorites, which tend to break up and flash during atmospheric entry. Furthermore since SQM is expected to be of interstellar origin, the speed of the track would be around 220 km/s (galactic velocity), higher than that of meteors that have an average speed of 40 km/s (solar system velocity), although interstellar meteors have also been observed \cite{interstellarmeteor}. 

Strangelets \cite{Madsen2004} may represent a fraction of UHECR. The first `evidence' of such form of SQM could be provided by Price's event (balloon) that was initially proposed as a magnetic monopole candidate and subsequently rebutted and interpreted instead as a strangelet, having penetrated the Earth atmosphere.

\subsection{Penrose process}

After gravitational collapse of a NS a (rotating) black hole (BH) can form that is the only astrophysical object that could carry out a tidal tearing of a Quark Star (QS) -- thus providing a source of SQM to be scattered throughout the Universe\footnote{One should however keep in mind the objections in \cite{Madsen2004}, since there is no impact with stellar protons and no disintegration of SQM in the process.}. 

Rotating (Kerr) BHs are special in that in addition to a horizon that hides the curvature singularity they are surrounded by an `ergo-region' external to the horizon where a time-like Killing vector becomes space-like \cite{KerrBH}. Although curvature is finite in the ergo-region, tidal forces are much stronger than outside the ergo-sphere and a compact object captured by the rotating BH may be torn into pieces with different energies and angular momenta. As a result Penrose process can take place in Kerr BHs. This will be reviewed in the Appendix for the reader's convenience and for comparison with the case of BPS fuzz balls.

According to Penrose \cite{Penrose:1971uk} a particle with positive energy (wrt to flat infinity) can enter the ergo-region and split into two (or more) particles one of which has negative energy (as seen from infinity) and crosses the horizon to finally fall into the singularity. The rest of the products can escape back to infinity carrying out more energy than their `mother'. The extra energy is provided by the BH that loses energy and angular momentum, since the initial particle should be counter-rotating, ie have opposite angular momentum wrt to the BH, for the very process to take place. The efficiency of the process is defined as the energy gained wrt to the initial energy
$$
\eta = \frac{{\cal E}_f - {\cal E}_i}{{\cal E}_i}
$$
As we will see momentarily, the efficiency depends on the mass, energy and spin of the initial particle as well as of the products and of the BH and the place where the splitting takes place. In the simple case when the initial counter-rotating object with energy equal to its rest mass (${\cal E}_0=\mu_0$) decomposes into two massless products at the turning point of its geodesics in the equatorial plane ($\theta = \pi/2$), it is simple to express $\eta$ in terms of the `radial' position $r^*$ where the process takes place. We have reproduced the text-book analysis in the Appendix for the interested reader and for comparison with the similar process in non-BPS fuzz balls. Though almost obvious $\eta$ is positive -- in fact with an upper bound $\eta\le (1/2)(\sqrt{2}-1)$ -- only when $r^*$ lies inside the ergo-region and the particle is massive and counter-rotating. The analogue process for massless particles or waves is called super-radiance \cite{SuperRad}. 

\subsection{Acceleration mechanisms for UHECR}

Thanks to Penrose mechanism Kerr/rotating  BHs can be used as  cosmic slings thus allowing one to reach the peak (GZK cutoff) of the UHECRs mountain. Sling-shot by other small magnetized objects such as white dwarves, neutron stars and quark stars has been proposed by Blandford \& Znajek, Berti, Brito \& Cardoso, Banados and West. Contrary to BHs the efficiency $\eta$ is not bounded from above. As we will momentarily see, this will turn out to be the case for non-BPS `fuzz-balls', too. 

Before concluding this section, let us briefly recall the broad features of the two classes of acceleration mechanisms for CR proposed so far. 

According to the first ``one-shot'' mechanism, CR are accelerated by an extended/intense electric field directly to the ZeV scale~\cite{Hillas:1985is}. The original idea put forward by Swann~\cite{Swann} has been elaborated on and the necessary electric field is usually related to the fast rotation of small, highly magnetized objects such as white dwarfs~\cite{deJager,Ikhsanov:2005qf}, neutron stars (pulsars)~\cite{Gunn:1969ej,Blasi:2000xm,Arons:2002yj, Fang:2012rx, Fang:2013cba}, or black holes~\cite{Blandford:1977ds, Znajek, Lovelace}. While electric field acceleration has the advantage of being fast, it suffers from the drawback of occurring in astrophysical sites with extremely high energy density, where many energy-loss phenomena can take place at the same time. 

According to the second ``stochastic'' mechanism  of acceleration, instead, particles gain energy gradually through multiple interactions with moving magnetized plasmas. The idea, pioneered by Fermi~\cite{Fermi:1949ee,Fermi:1954ofk}, can be realised in a variety of astrophysical environments, including the interplanetary medium~\cite{Jokipii:1971,Wenzel:1989}, supernova remnants (SNRs)~\cite{Scott:1975,Chevalier:1976,Chevalier:1978qk,Cowsik:1984yya,Torres:2002af,Blasi:2010gr}, the galactic disk and halo~\cite{Jokipii:1985, Jokipii:1987, Bustard:2016swa, Merten:2018qoa}, AGN's~\cite{Protheroe:1983,Kazanas:1985ud,Protheroe:1992qs}, large-scale jets and lobes of giant radio-galaxies (RG)~\cite{Biermann:1987ep,Rachen:1992pg,Romero:1995tn}, blazars~\cite{Blandford:1979za,Mannheim:1993jg,Dermer:2008cy,Caprioli:2015zka}, gamma-ray bursts (GRBs)~\cite{Waxman:1995vg,Vietri:1995hs}, starburst superwinds~\cite{Anchordoqui:1999cu, Anchordoqui:2018vji}, galactic microquasar systems~\cite{Levinson:2001as,Aharonian:2005cx}, and clusters of galaxies~\cite{Norman:1995,Kang:1996rp,Ryu:2003cd}. Contrary to the previous case, stochastic acceleration tends to be slow.  Furthermore it poses the issue of how to keep relativistic particles confined within the Fermi `engine'.

\section{Penrose mechanism for smooth non-BPS fuzzballs}

In this Section, we analyze the Penrose process for neutral massive scalar particles in smooth non-BPS geometries such as JMaRT. In order to set the stage for the computation we will first recall JMaRT soliton solution and its properties and then study the geodesics motion in this geometry. Thanks to the large amount of isometry the problem is integrable very much as for Kerr BHs\footnote{We thank P.~Fr{\'e} for stressing this property.} as well as for some BPS fuzz-balls \cite{D1D5fuzzgeod}. To exploit this property it is convenient to work in the Hamiltonian formulation that requires the determination of the canonical momenta $P_\mu$, conjugate to the generalised velocities $\dot{x}^\mu$. We will restrict our attention on the case where the conserved KK momentum $P_y$ of the infalling particle is zero. Moreover, we focus on geodesics in the hyperplane $\theta = 0$ whereby an effective dimensional reduction takes place since the radius of one of the angular directions ($\phi$) shrinks to zero and one has to set the corresponding conserved (angular) momentum $P_\phi$ to zero. One ends up with only three variables $t,r,\psi$ and the dynamics looks remarkably similar to the one in Kerr BHs\footnote{The same happens for $\theta = \pi/2$ after replacing $\psi$ with $\phi$ and the parameters $a_1$ and $a_2$ with one another.}. 

Despite the relatively compact and elegant form of JMaRT, explicit formulae for the `effective potentials' ${\cal E}_{\pm}^{\uparrow\uparrow/\uparrow\downarrow}$ and for the efficiency $\eta$ tend to become unwieldy. We will express the results in compact form in terms of the coefficient functions that appear as components of the inverse metric. We refrain from displaying cumbersome formulae that cannot illuminate the understanding. To illustrate the results for various values of the parameters we present different plots of ${\cal E}_{\pm}^{\uparrow\uparrow/\uparrow\downarrow}$ and $\eta$ as well as for other relevant coefficient functions. 

\subsection{JMaRT solution and its properties}

In string theory, the objects colloquially called black-holes (BHs) are bound states of strings and p-branes, {\it i.e.} p-dimensional extended solitons. This description allows reproducing the micro-states necessary to explain the origin of BH entropy that scales with the area of the event horizon, at least for charged BPS black-holes \cite{StromVafaetc}. In the fuzzball proposal \cite{FuzzMathur} classical BHs can be thought of as ensembles of smooth, horizon-less geometries with the same asymptotic behaviour as the would-be BH, {\it i.e.} same mass, charge and angular momenta. BPS systems with two charges give rise to small BHs with string-size horizon. In order to have a large BH with a finite (possibly large) area of the event horizon, one has to consider systems with at least three charges in $D=5$ or four charges in $D=4$. One of the grand successes of string theory is the precise micro-state counting for charged BHs, mostly in a BPS context. Extension to non-BPS and un-charged BHs has proven much harder. 

For our purposes, as a toy model of the Penrose mechanism for non-BPS fuzz-balls, we will consider a non-BPS 3-charge solution in $D=5$ originally found by Jejjalla, Madden, Ross and Titchener (JMaRT). JMaRT solutions\footnote{Henceforth we call it JMaRT for short.} in Type IIB superstring theory depend on five parameters associated to charges: D1-brane $Q_1$ and D5-brane $Q_5$ charge, the asymptotic radius $R$ of the Kaluza-Klein circle and two additional integer parameters $m$ and $n$. For $m=n{+}1$ the solutions turn out to be BPS. Imposing appropriate conditions on the parameters, that determine the mass and angular momenta, JMaRT has neither singularity nor event horizon and is free from CTC's. 

The reason why we are interested in JMaRT is the presence of an ergo-region, whereby particles with negative energy can propagate. It has been argued that an ergo-region that does not enclose a horizon and a singularity should lead to an instability: JMaRT should decay to an extremal BPS solution with the same charges. This ergo region or similar instabilities has been studied by various groups \cite{JMaRTinstab, BPSinstab}.
We will assume that the decay process would take a long time so much so that JMaRT could behave as a cosmic sling thanks to Penrose process, that in turn can also play a role in the relaxation of JMaRT to a stable BPS configuration. Having in mind SQM and strangelets, we focus on massive neutral scalar particles rather than on waves. The analogous process for waves is called `super-radiance' and has been studied for JMaRT in \cite{SuperRad}.  Hawking process has also been considered for JMaRT in \cite{MathurHawk}.

In order to construct JMaRT one starts fromType IIB supergravity in $D=10$ and considers 3-charge micro-state geometries of the D1-D5-P system \cite{GiustoMatSaxetc}. The D1-branes wrap a circle $S^1_y$, along which KK-momentum is added, while the D5-branes wrap a five-torus $S^1_y\times T^4$. The original solution depends on 8 parameters that determine the mass $M_{ADM}$ (related to $M$), two independent angular momenta $J_\phi$ and $J_\psi$ (parameterised in terms of $a_1$ and $a_2$), the three charges $Q_1$, $Q_5$ and $Q_{p}$ (expressible in terms of the `boost' parameters $\delta_1$, $\delta_5$ and $\delta_{p}$), the radius $R$ of the $S^1_y$ and the volume $V_4$ of the four-torus $T^4$. 

Safely neglecting $T^4$, whose volume can be taken to be very small, the six-dimensional geometry is parameterized in terms of  $t$ (time), $r$ (radial coordinate), $y$ (for $S^1_y$) and three  angular coordinates $\theta$, $\phi$ and $\psi$ and reads 
\begin{align}
&{{d}}s^2=
\frac{M ({s_{p}}{{{d}}y} -{c_{p}} {{{d}}t})^2}{\sqrt{{H_1} {H_5}}}
        -\frac{f \left({{{d}}t}^2-{{{d}}y}^2\right)}{\sqrt{{H_1} {H_5}}}
+\sqrt{{H_1} {H_5}} \left[\frac{r^2{{{d}}r}^2 }{\left(r^2+{a_1}^2\right) \left(r^2+{a_2}^2\right)-M r^2}+{{{d}}\theta }^2\right]\nonumber\\
&+{s_{\theta} }^2 {{{d}}\phi }^2 \left[\sqrt{{H_1} {H_5}}+\frac{\left({a_2}^2{-}{a_1}^2\right){s_{\theta} }^2 ({H_1}{+}{H_5}{-}f)}{\sqrt{{H_1} {H_5}}}\right] +
{c_{\theta} }^2 {{{d}}\psi }^2 \left[\sqrt{{H_1} {H_5}}+\frac{\left({a_1}^2{-}{a_2}^2\right){c_{\theta} }^2 ({H_1}{+}{H_5}{-}f)}{\sqrt{{H_1} {H_5}}}\right]
\nonumber \\ &+\frac{ 2 M {s_{\theta} }^2{{{d}}\phi }[{{{d}}t} ({a_2} {c_1} {c_5} {c_{p}}-{a_1} {s_1} {s_5} {s_{p}})+{{{d}}y} ({a_1} {c_{p}} {s_1} {s_5}-{a_2} {c_1} {c_5} {s_{p}})]}{\sqrt{{H_1} {H_5}}}
        +\frac{M \left({a_1} {c_{\theta} }^2 {{{d}}\psi }+{a_2} {s_{\theta} }^2{{{d}}\phi } \right)^2}{\sqrt{{H_1} {H_5}}} \nonumber
\\ &+\frac{2 M {c_{\theta} }^2 {{{d}}\psi } [{{{d}}t} ({a_1} {c_1} {c_5} {c_{p}}-{a_2} {s_1} {s_5} {s_{p}})+{{{d}}y} ({a_2} {c_{p}} {s_1} {s_5}-{a_1} {c_1} {c_5} {s_{p}})]}{\sqrt{{H_1} {H_5}}}
\end{align}
where\footnote{Our $H_i$ are denoted by ${\widetilde{H}}_i$ in JMaRT \cite{JMaRT}.}
\begin{eqnarray} 
{{H}}_{i}=f+M\sinh^2\delta_i \qquad , \qquad
f=r^2+a_1^2\sin^2\theta+a_2^2\cos^2\theta,
\end{eqnarray}
with $c_i = \cosh \delta_i$, $s_i = \sinh \delta_i$, for short henceforth, as well as $c_{\theta}=\cos \theta$, $s_{\theta}=\sin \theta$, $c_{\phi}=\cos \phi$, $s_{\phi}=\sin \phi$, $c_{\psi}=\cos \psi$, $s_{\psi}=\sin \psi$.  


We have not displayed the profiles of the other Type IIB supergravity fields that are present in JMaRT since they play no role in our later analysis of the Penrose process. 

The 6-dimensional metric can be written in the form
\begin{multline}
{{{d}}s}^2=-A\ {{{d}}t}^2+B\ {{{d}}r}^2+\ C_{\psi }{{{d}}\psi }^2+\ C_{\phi }{{{d}}\phi }^2+U{{{d}}\theta }^2 +F\ {{{d}}y}^2\\
+2 \Omega _{\psi }\ {{{d}}t} {{{d}}\psi } +2\Omega _{\phi }\ {{{d}}t} {{{d}}\phi +2K\ {{{d}}t} {{{d}}y}  } 
+2\Lambda _{\psi }\ {{{d}}y}{{{d}}\psi }  +2\Lambda _{\phi }\ {{{d}}y} {{{d}}\phi } +2 \Gamma\ {{d}}\psi {{d}}\phi
\end{multline}
with
\begin{equation}
    -A =\frac{-f+{c_{p}}^2 M}{\sqrt{{H_1} {H_5}}} \quad , \quad 
    B =\frac{r^2 \sqrt{{H_1} {H_5}}}{\left({a_1}^2+r^2\right) \left({a_2}^2+r^2\right)-M r^2} \quad , \quad 
    U =\sqrt{{H_1} {H_5}} \quad , \quad F =\frac{f+M {s_{p}}^2}{\sqrt{{H_1} {H_5}}} 
    \end{equation}\begin{equation}
    C_{\psi } =\frac{{a_1}^2 {c_{\theta} }^4 (-f+{H_1}+{H_5}+M)+{a_2}^2 {c_{\theta} }^4 (f-{H_1}-{H_5})+{c_{\theta} }^2 {H_1} {H_5}}{\sqrt{{H_1} {H_5}}} \end{equation}\begin{equation}
    C_{\phi } =\frac{{s_{\theta} }^4 \left(({a_1}-{a_2}) ({a_1}+{a_2}) (f-{H_1}-{H_5})+{a_2}^2 M\right)+{s_{\theta} }^2 {H_1} {H_5} }{\sqrt{{H_1} {H_5}}} \\
    \end{equation}
\begin{equation}
    \Gamma =\frac{{a_1} {a_2} {c_{\theta} }^2 M {s_{\theta} }^2}{\sqrt{{H_1} {H_5}}} \quad , \quad \Omega _{\psi } =\frac{{c_{\theta} }^2 M ({a_1} {c_1} {c_5} {c_{p}}-{a_2} {s_1} {s_5} {s_{p}})}{\sqrt{{H_1} {H_5}}} \quad , \quad
    \Omega _{\phi } =\frac{{s_{\theta} }^2 M ({a_2} {c_1} {c_5} {c_{p}}-{a_1} {s_1} {s_5} {s_{p}})}{\sqrt{{H_1} {H_5}}} 
     \\
  \end{equation}
 \begin{equation} 
    K =-\frac{{c_{p}}{s_{p}} M }{\sqrt{{H_1} {H_5}}} \quad , \quad
    \Lambda _{\psi } =\frac{{c_{\theta} }^2 M ({a_2} {c_{p}} {s_1} {s_5}-{a_1} {c_1} {c_5} {s_{p}})}{\sqrt{{H_1} {H_5}}} \quad , \quad
    \Lambda _{\phi } =\frac{{s_{\theta} }^2 M ({a_1} {c_{p}} {s_1} {s_5}-{a_2} {c_1} {c_5} {s_{p}})}{\sqrt{{H_1} {H_5}}} \end{equation}

 The ADM mass and angular momenta are given by 
 \begin{equation}
  M_{ADM} = \frac{M}{2} \sum_i \cosh{2\delta_i} 
  \quad , \quad J_\phi = M(a_1 s_1s_2s_p - a_2 c_1c_5c_p) \quad , \quad  J_\psi = M(a_2 s_1s_2s_p - a_1 c_1c_5c_p)
  \end{equation}
where $\delta_i\ge 0$, without loss of generality, and $c_i=\cosh\delta_i$ and $s_i=\sinh\delta_i$, as before. Note that $J_\phi$ and $J_\psi$ get exchanged under the exchange of $a_1$ and $a_2$.

Potential singularities appear when ${{H}}_1=0$ or $ {{H}}_5=0$ (curvature singularities) and when $\det{g}=0$, where
\begin{equation}
|\det{g}|= r^2 {{H}}_1 {{H}}_5\cos\theta^2 \sin\theta^2
\end{equation}
that is for $r^2=0$ (coordinate singularity) or for $\theta=0,\pi$ or $\theta = \pi/2$ (degeneration of the polar coordinates on the `poles' of $S^3$). The vanishing of $G(r) = (r^2+a_1^2)(r^2+a_2^2) - Mr^2$, the denominator of $g_{rr}$, at 
 \begin{equation}
 r^2_\pm = \frac{1}{2} \left[ (M-a_1^2 - a_2^2) \pm \sqrt{ (M-a_1^2 - a_2^2)^2 - 4 a_1^2a_2^2}\right]
  \end{equation}
require a detailed analysis. In order shows that $r=0$ is a removable coordinate singularity it proves convenient to introduce the adimensional variable 
\begin{equation}
x = {r^2-r_+^2\over r_+^2-r_-^2} \quad {\rm so \  that} \quad dx= {2rdr\over r_+^2-r_-^2}
\end{equation}
Moreover, if one could smoothly shrink a circle to zero at the origin ($x=0$), the space is capped at $x=0$ {\it i.e.} at $r^2 = r_+^2>r_-^2$ and the `true' curvature singularity at $x=-1$ {\it i.e.} at $r^2 = r_-^2$ is excised. 

Absence of singularities, horizons and closed-time-like curves imposes conditions on the parameters that can be satisfied in the low mass (parameter) regime 
\begin{equation}
M\le (a_1-a_2)^2
\end{equation}
and fixes $M$ and $R$ to be given by
\begin{equation}
M= a_1^2 + a_2^2 - a_1 a_2 \frac{c_1^2c_5^2 c_p^2 + s_1^2s_5^2 s_p^2}{c_1c_5c_p s_1s_5s_p} \qquad , \qquad R=\frac{M c_1c_5s_1s_5 \sqrt{c_1c_5c_p s_1s_5s_p}}{\sqrt{a_1a_2} (c_1^2c_5^2 c_p^2 - s_1^2s_5^2 s_p^2)}
\end{equation}
As a result one gets
\begin{equation}
r_-^2< r_+^2 = - a_1a_2 \frac{s_1s_5s_p}{c_1c_5c_p} < 0 
\end{equation}
Two quantization conditions (needed to have closed orbits for $\widetilde\phi = \phi +\alpha(s_i,c_i) y$ and $\widetilde\psi = \psi + \beta(s_i, c_i) y$ as $y\rightarrow y + 2\pi R$) constrain the remaining parameters in terms of two integers $m$ and $n$
\begin{equation}
\frac{j + j^{-1}}{s + s^{-1}}= m-n  \quad , \quad \frac{j - j^{-1}}{s - s^{-1}}= m+n
\end{equation}
where $j = \sqrt{a_2/a_1} \le 1$  and $s = \sqrt{ s_1s_5s_p/c_1c_5c_p} \le 1$, indeed one can take $a_1\ge a_2\ge 0$ without loss of generality,  thus getting $m\ge n{+}1\ge 1$. 
In terms of $j,s$ and $a_1$ the expression for $M$ reads 
\begin{equation}
    M(j,s)= {a_1}^2 \left({j}^4-{j}^2 {s}^2-\frac{{j}^2}{{s}^2}+1\right)
\end{equation}
replacing $j,s$ in terms of the integers $m,n$ one finds
\begin{align*}
    M(m,n) &= \frac{{a_1}^2 }{2 m^2 n^2}[m^2-(n+1)^2][m^2-(n-1)^2] \\
    &\{(m^2-n^2)^2 - (m^2-n^2)\sqrt{[m^2-(n+1)^2][m^2-(n-1)^2]}-m^2-n^2\}\numberthis
\end{align*}
that vanishes in the BPS case $m=n+1$ whereby $M\rightarrow 0$, $\delta_i\rightarrow \infty$ with $Q_i =M s_ic_i$ fixed. 

The remaining five independent parameters correspond to $Q_1$, $Q_5$, $R$, $m$ and $n$ that determine the KK charge $Q_p$ and the angular momenta $J_\phi$, $J_\psi$ 
\begin{equation}
Q_{p}  = nm\frac{Q_1Q_5}{R^2}\qquad , \qquad J_\phi = - m\frac{Q_1Q_5}{R} \qquad , \qquad J_\psi = n\frac{Q_1Q_5}{R}
\end{equation}

The Penrose process can take place in JMaRT thanks to the presence of an ergoregion, that can be identified as the region where the norm of the time-like Killing vector $V_t = \partial_t$ becomes positive. Using JMaRT one finds 
\begin{equation} 
||V_t||^2 = g_{\mu\nu} V^\mu_t V^\nu_t = g_{tt} = {Mc_p^2 - f \over \sqrt{{{H}}_1{{H}}_5}}
\end{equation} 
where $f(r,\theta) = r^2 + a_1^2\sin\theta^2 + a_2^2\cos\theta^2$ and ${{H}}_i = f(r,\theta) + M s_i^2$. 
An ergo-sphere appears at $f(r,\theta) = Mc_p^2$ 
\begin{equation} 
r_e^2 = Mc_p^2 - a_1^2\sin\theta^2 - a_2^2\cos\theta^2 
\end{equation} 
where $V_t$ becomes space-like. In the BPS limit the norm of $V_t$ is always negative: $||V_t||^2 = {- f /\sqrt{{{H}}_1{{H}}_5}}$ and no ergo-region appears.

\subsection{Geodetic motion in JMaRT}

As a preliminary step to investigate the Penrose process in JMaRT, we study the geodesics for massive or massless neutral particles. Probes of this kind only feel the presence of the curved metric but are unaffected by the other Type IIB fields present in JMaRT. 

The Lagrangian that governs geodetic motion is given by 
\begin{equation}
    \mathcal{L}=\frac{1}{2}g_{\mu\nu}\dot x^{\mu}\dot x^{\nu}
\end{equation}
where $g_{\mu\nu}$ denotes the six-dimensional metric tensor\footnote{The extra four directions compactified on $T^4$ play no role in our analysis.}. Recall that $dr$ and $d\theta$ appear diagonally in $ds^2$, while $dt, dy, d\phi, d\psi$  form a four-dimensional block. As in Kerr BH or in BPS fuzz balls, in order to take advantage of all the symmetries, {\it i.e.} time translation, KK shifts $U(1)_y$ and rotations $U(1)_\phi \times U(1)_\psi$, it is better to switch to the Hamiltonian formalism. The generalized momenta are given by
\begin{equation}
P_{\mu}=\frac{\partial \mathcal{L}}{\partial \dot x^\mu}= g_{\mu\nu}\dot x^{\nu}
\end{equation}
where $\dot x^{\nu} = dx^\mu/d\tau$ and the Hamiltonian reads
\begin{equation}
\mathcal{H}=P_{\mu}\dot x^{\mu} -\mathcal{L}=\frac{1}{2}g^{\mu\nu}P_{\mu}P_{\nu}
\end{equation}
where $g^{\mu\nu}$ is the inverse six-dimensional metric. 
For JMaRT the explicit expressions for the generalized momenta read 
\begin{align}
        P_r= 
        \frac{ {\dot r} r^2\sqrt{{H_1} {H_5}}}{\left({a_1}^2+r^2\right) \left({a_2}^2+r^2\right)-M r^2}
\end{align}
\begin{align}
        P_{\theta}={\dot \theta} \sqrt{{H_1} {H_5}}
\end{align}
\begin{align}
        P_t = -\frac{{\dot t} \left(f-{c_{p}}^2 M\right)}{\sqrt{{H_1} {H_5}}}-\frac{{\dot y}{c_{p}} M {s_{p}} }{\sqrt{{H_1} {H_5}}}+
        \frac{{\dot \psi} {c_{\theta}}^2 M ({a_1} {c_1} {c_5} {c_{p}}-{a_2} {s_1} {s_5} {s_{p}})}{\sqrt{{H_1} {H_5}}}
        +\frac{ {\dot \phi} M {s_{\theta}}^2({a_2} {c_1} {c_5} {c_{p}}-{a_1} {s_1} {s_5} {s_{p}})}{\sqrt{{H_1} {H_5}}}      
\end{align}
\begin{align}
        P_y=\frac{{\dot y} \left(f+M {s_{p}}^2\right)}{\sqrt{{H_1} {H_5}}}-\frac{{\dot t}{c_{p}} M {s_{p}} }{\sqrt{{H_1} {H_5}}}+
      \frac{{\dot \psi} {c_{\theta} }^2 M ({a_2} {c_{p}} {s_1} {s_5}-{a_1} {c_1} {c_5} {s_{p}})}{\sqrt{{H_1} {H_5}}}
        +\frac{{\dot \phi} M {s_{\theta} }^2 ({a_1} {c_{p}} {s_1} {s_5}-{a_2} {c_1} {c_5} {s_{p}})}{\sqrt{{H_1} {H_5}}}
\end{align}
\begin{align}
        P_{\phi}=&
        \frac{{\dot \phi} {s_\theta }^2\{{s_\theta }^2[({a^2_1}-{a^2_2})(f-{H_1}-{H_5})+{a^2_2} M] + {H_1} {H_5}\}}{\sqrt{{H_1} {H_5}}}
        \nonumber \\
        &+\frac{{\dot t} {s_\theta }^2 ({a_2} {c_1} {c_5} {c_{p}}-{a_1} {s_1} {s_5} {s_{p}})M}{\sqrt{{H_1} {H_5}}} +\frac{ {\dot y} {s_\theta }^2({a_1} {c_{p}} {s_1} {s_5}-{a_2} {c_1} {c_5} {s_{p}})M}{\sqrt{{H_1} {H_5}}}
        +\frac{{\dot \psi}{s_\theta }^2{c_\theta }^2 {a_1} {a_2}M   }{\sqrt{{H_1} {H_5}}}
\end{align}
\begin{align}
        P_{\psi}=&
        \frac{{\dot \psi} {c_\theta }^2\{{c_\theta }^2[({a^2_2}-{a^2_1})(f-{H_1}-{H_5})+{a^2_1} M] + {H_1} {H_5}\}}{\sqrt{{H_1} {H_5}}}
        \nonumber \\
        &+\frac{{\dot t} {c_\theta }^2 ({a_1} {c_1} {c_5} {c_{p}}-{a_2} {s_1} {s_5} {s_{p}})M}{\sqrt{{H_1} {H_5}}} +\frac{ {\dot y} {c_\theta }^2({a_2} {c_{p}} {s_1} {s_5}-{a_1} {c_1} {c_5} {s_{p}})M}{\sqrt{{H_1} {H_5}}}
        +\frac{{\dot \phi}{c_\theta }^2{s_\theta }^2 {a_1} {a_2}M   }{\sqrt{{H_1} {H_5}}}    
\end{align}

and the Hamiltonian for JMaRT can be written as
\begin{align}
H &= {1\over 2} \left\{- \widetilde{A} P_t^2 + \widetilde{B}P_r ^2 + \widetilde{U} P_\theta^2
+\widetilde{C}_\psi P_\psi^2 +\widetilde{C}_\phi P_\phi^2 +\widetilde{F} P_y^2 \right\} 
\nonumber\\
&+{P_t} {P_y} \widetilde{{K}} +{P_t} {P_{\psi}} \widetilde{\Omega}_{\psi}+{P_t} {P_{\phi}} \widetilde{\Omega}_{\phi}+{P_y}{P_{\psi}} \widetilde{\Lambda}_{\psi} +{P_y}{P_{\phi}} \widetilde{\Lambda}_{\phi}+{P_{\psi}} {P_{\phi}} \widetilde{\Gamma } 
\end{align}
where the coefficient functions $\widetilde{A}$, $\widetilde{B}$, $\widetilde{U}$, $\widetilde{C}_{\phi}$, $\widetilde{C}_{\psi}$, $\widetilde{F}$, $\widetilde{\Omega} _{\phi}$, $\widetilde{\Omega} _{\psi}$, $\widetilde{\Gamma}$, $\widetilde{K}$, $\widetilde{\Lambda}_{\phi}$, $\widetilde{\Lambda} _{\psi}$ are the non-zero components of the inverse metric $g^{\mu\nu}$, whose explicit expressions are quite cumbersome and will not be displayed, except for the special case of $\theta=0$.

The generalized velocities can be expressed in terms of the momenta using the above functions \begin{align}
   \dot t &= -{P_t} \widetilde{A}+{P_y} \widetilde{{K}}+{P_{\psi}} \widetilde{\Omega}_{\psi}+{P_{\phi}} \widetilde{\Omega}_{\phi} \\
    \dot r &= P_r  \widetilde{B} \\
    \dot \theta &= {P_{\theta}} \widetilde{U} \\
    \dot \phi &= {P_{\phi}} \widetilde{C}_{\phi}+{P_t} \widetilde{\Omega}_{\phi}+{P_{\psi}} \widetilde{\Gamma }+{P_y} \widetilde{\Lambda}_{\phi} \\
    \dot \psi &= {P_{\psi}} \widetilde{C}_{\psi}+{P_t} \widetilde{\Omega}_{\psi}+{P_y} \widetilde{\Lambda}_{\psi}+{P_{\phi}} \widetilde{\Gamma } \\
     \dot y &= {P_y} \widetilde{F}+{P_t} \widetilde{{K}}+{P_{\psi}} \widetilde{\Lambda}_{\psi}+{P_{\phi}} \widetilde{\Lambda}_{\phi}
\end{align}

Very much as for Kerr BH and for BPS fuzz-balls, the system is integrable in that the dynamics in the $r$ and $\theta$ coordinates can be  separated in principle. In practice the geodesics are non-planar and their explicit form is not very illuminating for our purposes.

Following similar analysis in BPS fuzz-balls \cite{D1D5fuzzgeod} and without losing any significant feature of the result, one can focus on the hyper-planes $\theta=0$ and $\theta = \pi/2$.  Indeed it is consistent to set $\dot\theta=0$ and $P_\theta = 0$ in these two cases since
\begin{equation}
{dP_{\theta}\over d\tau} = - {\partial {\cal H}\over \partial\theta} = 0 \qquad  {\rm for} \qquad  \theta=0 \quad {\rm and} \quad \theta = \pi/2 
\end{equation}
For both choices an effective dimensional reduction takes place. For $\theta=0$ all terms in $d\phi$ drop, being proportional to $\sin\theta^2$, and one can safely set $P_\phi=0$; while for $\theta=\pi/2$ all terms in $d\psi$ drop, being proportional to $\cos\theta^2$, and one can safely set $P_\psi=0$. The two cases are perfectly equivalent and one can get one from the other by simply exchanging $a_1$ and $a_2$ in any relevant formula. For definiteness we will focus on the $\theta = 0$ hyperplane in the following.

Moreover, we are not interested in the motion along the compact $y$ direction. In order to simplify the analysis, one can set $P_y = 0$. This is consistent since $P_y = 0$ is conserved: $\dot{P}_y = 0$. As a consequence $\dot{y}$ is completely determined by the other velocities and conserved momenta, so much so that we will not consider it later on. 



If we fix $\theta=0$ and consequently $P_{\theta}=0$, all the terms in $d\phi$ drop and the metric becomes

\begin{align}
&{{d}}s^2_{\theta=0}=\frac{{{{d}}t}^2 \left({c_{p}}^2 M-f\right)}{\sqrt{{H_1} {H_5}}}
        +\frac{{{{d}}r}^2 r^2 \sqrt{{H_1} {H_5}}}{\left({a_1}^2+r^2\right) \left({a_2}^2+r^2\right)-M r^2}
        +\frac{ {{{d}}\psi }^2 [ ({a_1}^2{-}a_2^2) ({H_1}{+}{H_5}{-}f)+{a_1}^2M+{H_1} {H_5}]}{\sqrt{{H_1} {H_5}}} \nonumber\\
&+{{{d}}t} \left(\frac{2 {{{d}}\psi } M ({a_1} {c_1} {c_5} {c_{p}}-{a_2} {s_1} {s_5} {s_{p}})}{\sqrt{{H_1} {H_5}}}-\frac{2 {c_{p}} {{{d}}y} M {s_{p}}}{\sqrt{{H_1} {H_5}}}\right)
        +\frac{2 {{{d}}\psi } {{{d}}y} M ({a_2} {c_{p}} {s_1} {s_5}-{a_1} {c_1} {c_5} {s_{p}})}{\sqrt{{H_1} {H_5}}}  +\frac{{{{d}}y}^2 \left(f+M {s_{p}}^2\right)}{\sqrt{{H_1} {H_5}}}      
\end{align}

which takes the form

\begin{equation}
{{{d}}s}^2_{\theta=0} = -A {{{d}}t}^2+B {{{d}}r}^2+C_{\psi }{{{d}}\psi }^2 +2\Omega_{\psi } {{{d}}t} {{{d}}\psi } +2 K {{{d}}t} {{{d}}y} +F {{{d}}y}^2 +2\Lambda_{\psi } {{{d}}\psi } {{{d}}y} = \hat{g}_{\mu\nu} dx^\mu dx^\nu
\end{equation}
where
\begin{align}    
-A&=\frac{{c_{p}}^2 M-f}{\sqrt{{H_1} {H_5}}}
\\
B&=\frac{r^2 \sqrt{{H_1} {H_5}}}{\left({a_1}^2+r^2\right) \left({a_2}^2+r^2\right)-M r^2}
\\
C_{\psi }&=\frac{{a_1}^2 (-f+{H_1}+{H_5}+M)+{a_2}^2 (f-{H_1}-{H_5})+{H_1} {H_5}}{\sqrt{{H_1} {H_5}}}
\\
\Omega_{\psi }&=\frac{M ({a_1} {c_1} {c_5} {c_{p}}-{a_2} {s_1} {s_5} {s_{p}})}{\sqrt{{H_1} {H_5}}}
\\
F&=\frac{f+M {s_{p}}^2}{\sqrt{{H_1} {H_5}}}
\\
K&=-\frac{{c_{p}} M {s_{p}}}{\sqrt{{H_1} {H_5}}}
\\
\Lambda_{\psi }&=\frac{M ({a_2} {c_{p}} {s_1} {s_5}-{a_1} {c_1} {c_5} {s_{p}})}{\sqrt{{H_1} {H_5}}}
\end{align}
with 
\begin{equation} f = r ^ { 2 } +{ a}_2^{ 2} \quad , \quad H _ { 1 } = f + M {s _ { 1 } }^{ 2 } \quad , \quad  H _ { 5 } = f + M {s _ { 5 }} ^ { 2 } 
\end{equation}
Recall that $a_1$ and $a_2$ switch their role under a change of $\theta$ from $0$ to $\pi/2$ and an exchange $\psi\leftrightarrow \phi$.

The reduced Hamiltonian expressed in terms of the components of the reduced inverse metric reads\begin{align}
    {\cal H} =\frac{1}{2} \left(-{P_t}^2 \widetilde{A}+P_r ^2 \widetilde{B}+{P_\psi }^2 \widetilde{C}_{\psi}+{P_y}^2 \widetilde{F}\right) + {P_t} {P_y} \widetilde{{K}}+ {P_t} {P_\psi } \widetilde{\Omega}_{\psi}+ {P_\psi } {P_y} \widetilde{\Lambda}_{\psi} = - \frac{\mu^2}{2}
\end{align}
where $\mu$ is the mass of the probe and the coefficient functions $\widetilde{A}$, $\widetilde{B}$, ${\widetilde{C}_{\psi }}$, $\widetilde{F}$, $\widetilde{\Omega}_{\psi}$, $\widetilde{K}$, ${\widetilde{\Lambda}_{\psi }}$ are the non-zero components of the reduced inverse metric and $\mu$ is the rest mass of the probe spin-less particle. 
At $\theta=0$ one has
\begin{equation}
\widetilde{B}=\frac{1}{B} = \frac{\left({a_1}^2+r^2\right) \left({a_2}^2+r^2\right)-M r^2}{r^2 \sqrt{{H_1} {H_5}}}
\end{equation} 
and
\begin{align*}
    \widetilde{A} &=\Bigl\{\sqrt{{H_1} {H_5}} \Bigl[{a_1}^2 \left(M {s_P}^2 \left(-{c_1}^2 {c_5}^2 M+{H_1}+{H_5}+M\right)-f^2+f \left({H_1}+{H_5}-M {s_P}^2+M\right)\right)\\
    &+2 {a_1} {a_2} {c_1} {c_5} {c_P} M^2 {s_1} {s_5} {s_P} +{a_2}^2 \Bigl(f^2-f \left({H_1}+{H_5}-M {s_P}^2\right) \\
    &-M \left({s_P}^2 ({H_1}+{H_5})+M {s_1}^2 {s_5}^2 \left({s_P}^2+1\right)\right)\Bigr)+{H_1} {H_5} \left(f+M {s_P}^2\right)\Bigr]\Bigr\}\\
    &\Bigl\{(f-M) \left({a_2}^2 \left(f^2-f ({H_1}+{H_5})-M^2 {s_1}^2 {s_5}^2\right)+f {H_1} {H_5}\right)\\
    &-{a_1}^2 f \left(M \left(-{c_1}^2 {c_5}^2 M+{H_1}+{H_5}+M\right)+f^2-f ({H_1}+{H_5}+2 M)\right)\Bigr\}^{-1} \numberthis
\end{align*}

\begin{align*}
    \widetilde{C}_{\psi} &=f (f-M) \sqrt{{H_1} {H_5}}\ \Bigl\{(f-M) \left({a_2}^2 \left(f^2-f ({H_1}+{H_5})-M^2 {s_1}^2 {s_5}^2\right)+f {H_1} {H_5}\right) \\
    &-{a_1}^2 f \left(M \left(-{c_1}^2 {c_5}^2 M+{H_1}+{H_5}+M\right)+f^2-f ({H_1}+{H_5}+2 M)\right)\Bigr\}^{-1} \numberthis
\end{align*}

\begin{align*}
   \widetilde{F} &= \sqrt{{H_1} {H_5}} \Bigl\{{a_1}^2 \Bigl[-\Bigl(M \left({s_P}^2+1\right) \left(-{c_1}^2 {c_5}^2 M+{H_1}+{H_5}+M\right)+f^2 \\
   &-f \left({H_1}+{H_5}+M \left({s_P}^2+2\right)\ \right)\Bigr)\Bigr]\\
   & -2 {a_1} {a_2} {c_1} {c_5} {c_P} M^2 {s_1} {s_5} {s_P}+{a_2}^2 \Bigl[f^2-f \left({H_1}+{H_5}+M {s_P}^2+M\right)\\
   & +M \left({H_1} {s_P}^2+{H_1}+{H_5} {s_P}^2+{H_5}+M {s_1}^2 {s_5}^2 {s_P}^2\right)\Bigr] +{H_1} {H_5} \left(f-M \left({s_P}^2+1\right)\right)\Bigr\} \\
   & \Bigl\{{a_1}^2 f \Bigl[M \left(-{c_1}^2 {c_5}^2 M+{H_1}+{H_5}+M\right)+f^2-f ({H_1}+{H_5}+2 M)\Bigr] \\
   & -(f-M) \Bigl[{a_2}^2 \left(f^2-f ({H_1}+{H_5})-M^2 {s_1}^2 {s_5}^2\right)+f {H_1} {H_5}\Bigr]\Bigr\}^{-1} \numberthis
\end{align*}

\begin{align*}
    \widetilde{\Omega}_{\psi} &= \Bigl\{M \sqrt{{H_1} {H_5}} ({a_1} {c_1} {c_5} {c_P} f+{a_2} {s_1} {s_5} {s_P} (M-f))\Bigr\} \\
    & \Bigl\{{a_1}^2 f \Bigl[M \left(-{c_1}^2 {c_5}^2 M+{H_1}+{H_5}+M\right)+f^2 -f ({H_1}+{H_5}+2 M)\Bigr] \\
    & -(f-M) \Bigl[{a_2}^2 \left(f^2-f ({H_1}+{H_5})-M^2 {s_1}^2 {s_5}^2\right)+f {H_1} {H_5}\Bigr]\Bigr\}^{-1} \numberthis
\end{align*}

\begin{align*}
    \widetilde{K} &= \Bigl\{M \sqrt{{H_1} {H_5}} \Bigl[{a_1}^2 {c_P} {s_P} \left(-{c_1}^2 {c_5}^2 M-f+{H_1}+{H_5}+M\right) \\
    & +{a_1} {a_2} {c_1} {c_5} M {s_1} {s_5} \left(2 {s_P}^2+1\right)+{c_P} {s_P} \left({a_2}^2 \left(f-{H_1}-{H_5}-M {s_1}^2 {s_5}^2\right)+{H_1} {H_5}\right)\Bigr]\Bigr\} \\
    & \Bigl\{{a_1}^2 f \Bigl[M \left(-{c_1}^2 {c_5}^2 M+{H_1}+{H_5}+M\right)+f^2-f ({H_1}+{H_5}+2 M)\Bigr] \\
    & -(f-M) \Bigl[{a_2}^2 \left(f^2-f ({H_1}+{H_5})-M^2 {s_1}^2 {s_5}^2\right)+f {H_1} {H_5}\Bigr]\Bigr\}^{-1} \numberthis
\end{align*}

\begin{align*}
    \widetilde{\Lambda}_{\psi} &= \Bigl\{M \sqrt{{H_1} {H_5}} ({a_1} {c_1} {c_5} f {s_P}+{a_2} {c_P} {s_1} {s_5} (M-f))\Bigr\} \\
    & \Bigl\{{a_1}^2 f \left(M \left(-{c_1}^2 {c_5}^2 M+{H_1}+{H_5}+M\right)+f^2-f ({H_1}+{H_5}+2 M)\right) \\
    & -(f-M) \left({a_2}^2 \left(f^2-f ({H_1}+{H_5})-M^2 {s_1}^2 {s_5}^2\right)+f {H_1} {H_5}\right)\Bigr\}^{-1} \numberthis
\end{align*}
Since we are not interested in motion along the compact circle direction we can safely set  
\begin{equation}
P_y=0
\end{equation}
and, for convenience of the notation,
\begin{equation}
P_t = - {\cal E} \quad , \quad P_{\psi}= {\cal J}
\end{equation}
so that we get
\begin{equation}
P_r ^2=\frac{1}{\widetilde{B}} \left({\cal E}^2 \widetilde{A}- {\cal J}^2 \widetilde{{C}}_{\psi} -2 {\cal E} {\cal J} \widetilde{\Omega}_{\psi}- \mu ^2\right) =\frac{\widetilde{A}}{\widetilde{B}}({\cal E}-{\cal E}_+)({\cal E}-{\cal E}_-) \ge 0
\end{equation}
where the `effective potentials' read
\begin{equation}
{\cal E}_\pm = \frac{{\cal J}\widetilde{\Omega}_{\psi}\pm\sqrt{{\cal J}^2 (\widetilde{\Omega}_{\psi}^2+\widetilde{A} \widetilde{{C}}_{\psi}) +\mu ^2 \widetilde{A}}}{\widetilde{A}}
\end{equation}

Since ${\widetilde{A}/\widetilde{B}}\ge 0$ always, one has either 
${\cal E}>{\cal E}_+>{\cal E}_-$ or ${\cal E}<{\cal E}_-<{\cal E}_+$
where, depending on whether the particle is co-rotating ($\uparrow\uparrow$), ie ${\cal J}J_{\psi}>0$, or 
counter-rotating ($\uparrow\downarrow$), ie ${\cal J} J_{\psi}<0$, one has  
$$
{\cal E}_+^{\uparrow\uparrow} \ge \mu  \quad , \quad {\cal E}_-^{\uparrow\uparrow} \le -\mu
$$
or
$$
{\cal E}_+^{\uparrow\downarrow} \le 0 \quad {\rm for} \quad r\le r_e \quad , \quad  {\cal E}_-^{\uparrow\downarrow} \le -\mu  
$$

As we will see momentarily, the Penrose process can only take place in the latter case.

\begin{figure}
    \centering
    \includegraphics[width=0.96\textwidth]{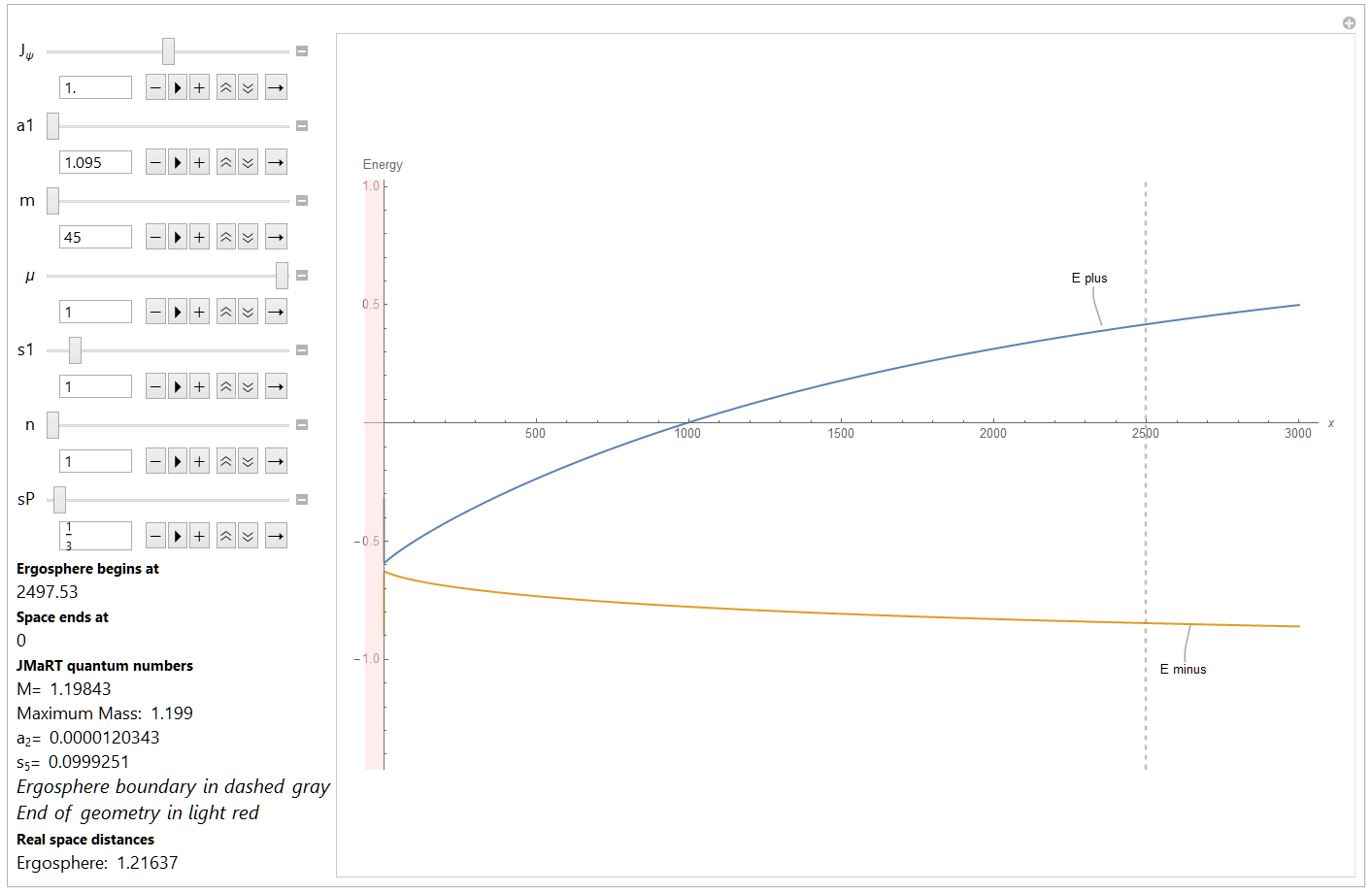}
    \caption{$ \cal E_+ ,\cal E_-$ regions for JMaRT with ``quantum numbers" shown on the left and a counter-rotating probe; $x = (r^2- {r_+}^2)/({r_+}^2-{r_-}^2)$. Ergosphere boundary is denoted by the dashed (grey online) vertical line on the right, end of geometry is in $x=0$.}
    \label{fig:NER1}
\end{figure}

\begin{figure}
    \centering
    \includegraphics[width=0.96\textwidth]{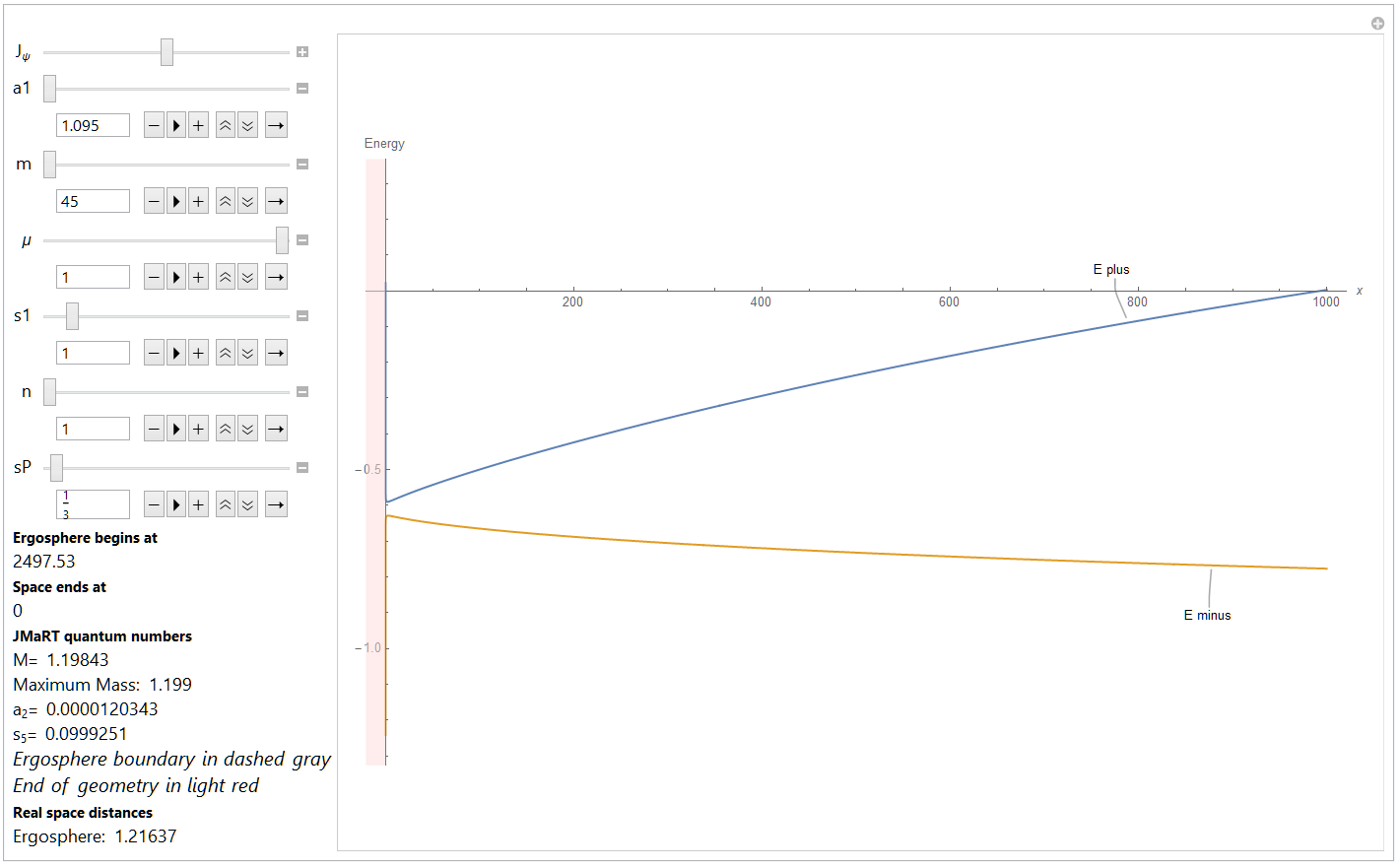}
    \caption{A close-up of the negative energy region for the same quantum numbers in Fig.\ref{fig:NER1}. Note the infinite wall exploding in the region close to $x=0$.}
    \label{fig:NER2}
\end{figure}

\begin{figure}
    \centering
    \includegraphics[width=\textwidth]{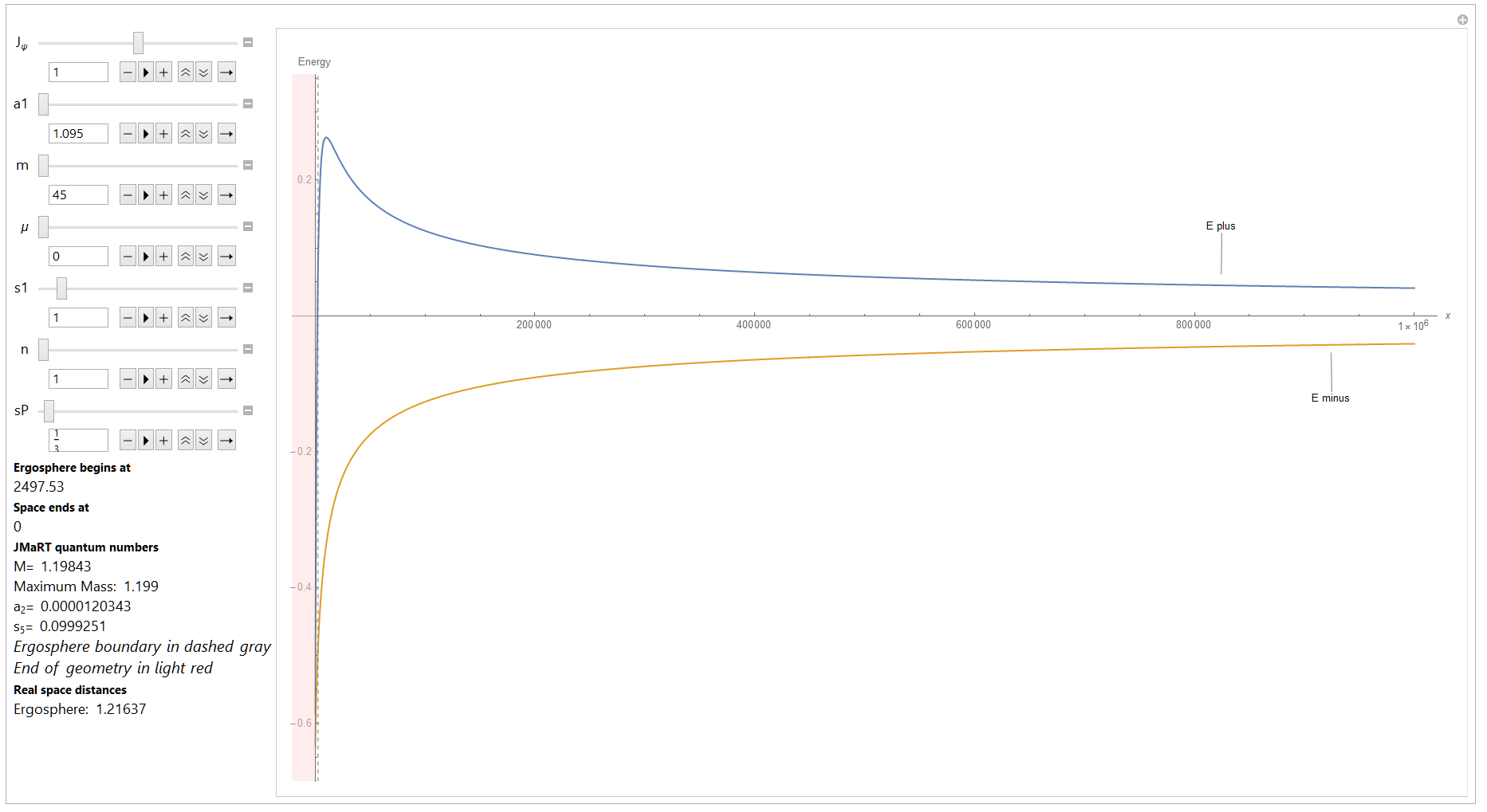}
    \caption{Asymptotic behaviour for the same quantum numbers as in Figs.\ref{fig:NER1}, \ref{fig:NER2}, but with probe mass $\mu=0$.}
    \label{fig:NER3}
\end{figure}

\begin{figure}
    \centering
    \includegraphics[width=\textwidth]{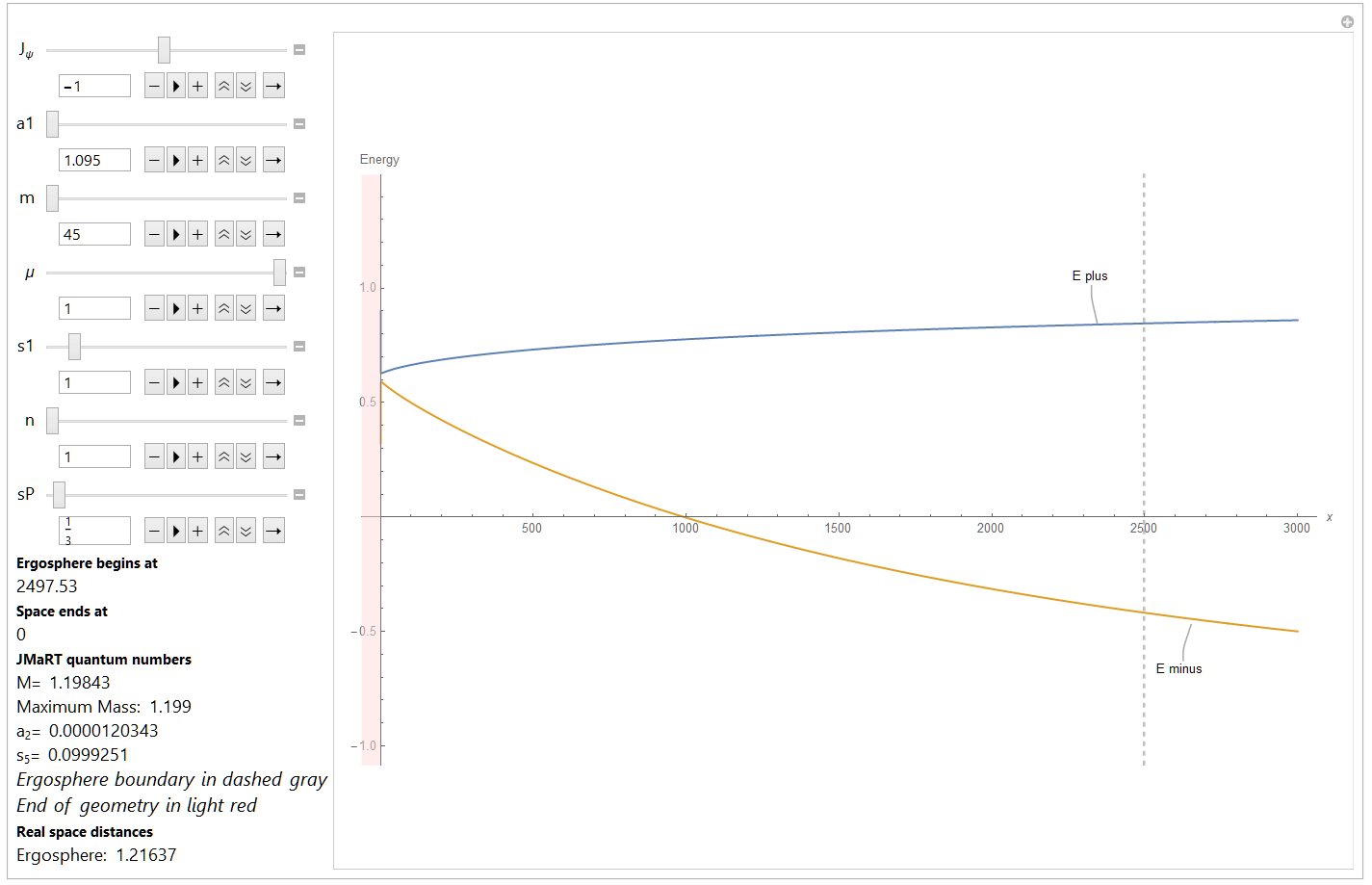}
    \caption{$ \cal E_+ ,\cal E_-$ regions for JMaRT with ``quantum numbers" shown on the left and a co-rotating probe; $x = (r^2- {r_+}^2)/({r_+}^2-{r_-}^2)$. Ergosphere boundary is denoted by the dashed (grey online) vertical line on the right, end of geometry is in $x=0$. There is no ergoregion.}
    \label{fig:NER4}
\end{figure}

In Figs. \ref{fig:NER1}, \ref{fig:NER2}, \ref{fig:NER3} we have plotted ${\cal E}^{\uparrow\downarrow}_\pm$ as a function of $r$ for some `reasonable' choice of the parameters $M, a_1, a_2$ of JMaRT (determined by the choice of $m,n,\delta_1,\delta_5$ and $a_1$ or, equivalently, $R$) and of the angular momentum ${\cal J}$ of the (massive $\mu\neq 0$ or mass-less $\mu= 0$) probe particle. Despite their complexity, thanks to the existence of a frame-dragging term ${{d}}t{{d}}\psi$ in Eq. \ref{eq:metricABC}, these solutions expose the expected presence of regions with negative energy inside the ergo-sphere that can be studied computationally and plotted. For comparison we also plot ${\cal E}^{\uparrow\uparrow}_\pm$ in Fig. \ref{fig:NER4}.

As evident from the plots of ${\cal E}_\pm$ there are two kinds of geodesics in the $\theta = 0$ plane: unbounded ones for ${\cal E}\ge \mu$ and trapped ones ${\cal E}\le - \mu$. In the former case the massive probe impinges from infinity, reaches a turning point $r^*$ where $P_r=0$ and gets deflected back to infinity, possibly after making several turns around the `center' ($x=0$ ie $r^2=r_+^2<0$) of the fuzzball. In the latter case, the particle cannot escape to infinity and remains forever inside the fuzzball. The relevant equations can be integrated in terms of non-elementary functions and we will not attempt to present a detailed analysis here. Instead we turn our attention on the Penrose process in JMaRT.  

\subsection{Penrose process in JMaRT and its efficiency}

The presence of an ergo-region in JMaRT allows the Penrose process to take place, whereby a counter-rotating particle acquires negative energy after crossing the ergo-sphere and if it splits into two or more fragments, one of the product may escape to infinity with an energy larger than the initial particle, while the other fragment(s) get trapped in the fuzzball for a `long' time. 

Following \cite{Chandra, Shutz}, {\it mutatis mutandis} we will derive the efficiency of Penrose process in  JMaRT metric. 

Let us consider a spin-less probe with rest mass $\mu_0$, energy  ${\cal E}_0\ge \mu_0$ (positive branch) and orbital angular momentum ${\cal J}_0$ opposite to the angular momentum of JMaRT (counter-rotating $\uparrow\downarrow$). Very much as in Penrose original analysis, it seems reasonable and computationally convenient 
to assume that the probe splits exactly at the turning point $r=r^*$ where $P_r(r^*)=0$. At this point the angular velocity reaches a maximum and the tidal tearing of the probe is more likely to take place. This has the additional advantage of simplifying the analysis since one gets a relation between $r$, ${\cal E}_0$ and ${\cal J}_0$ of the form
\begin{equation}
{\cal E}_0^2 \widetilde{A} -{\cal J}_0^2 \widetilde{C}_{\psi}-2{\cal E}_0{\cal J}_0\widetilde{\Omega}_{\psi}=\mu_0^2
\end{equation}
which can be (implicitly) solved for $r=r^*$ as a function of ${\cal E}_0$ (positive counter-rotating branch) and ${\cal J}_0$.

Denoting by ${\cal E}_1$, ${\cal E}_2$ and ${\cal J}_1$, ${\cal J}_2$ the energies and (orbital) angular momenta of the two (spin-less) fragments with rest masses $\mu_1$ and $\mu_2$, conservation of energy and angular momentum yield
\begin{equation}
    \left\{ \begin{array} { l } { \vphantom{\dfrac{1}{1}} {\cal E}_1+{\cal E}_2 = {\cal E}_0} \\ { \vphantom{\dfrac{1}{1}}  {\cal J}_1+{\cal J}_2 = {\cal J}_0} \end{array} \right.
\end{equation}
which has two solutions, symmetric under the exchange $1\leftrightarrow 2$. Assuming that particle 1 escapes to infinity (positive branch of the energy) while particle 2 gets trapped in the fuzzball (negative branch of the energy), one can first express ${\cal J}_1$ and ${\cal J}_2$ in terms of ${\cal E}_1={\cal E}_+(\mu_1)$ and ${\cal E}_2= {\cal E}_-(\mu_2)$ and get\footnote{We are implicitly assuming that the fragments are produced with zero radial momentum and continue to move in the $\theta=0$ plane with $P_y=0$ and $P_\phi=0$. This means that $r^*$ is a turning point for the fragments, too.} 
\begin{equation}
{\cal J}_{1,2}=\frac{-{\cal E}_{1,2}\widetilde{\Omega}_{\psi}\pm\sqrt{{\cal E}_{1,2}^2 (\widetilde{\Omega}_{\psi}^2 +\widetilde{A} \widetilde{{C}}_{\psi}) -\mu_{1,2}^2 \widetilde{{C}}_{\psi}}}{\widetilde{{C}}_{\psi} }
\end{equation}
Plugging these in the second equation and solving the system for ${\cal E}_{1,2}$ yields
\begin{equation}
{\cal E}_{1,2} =\frac{1}{2 {\mu_0}^2}\left\{{\cal E}_{0} ({\mu_0}^2\pm{\mu_1}^2\mp{\mu_2}^2)
\pm\sqrt{{\cal F}({\mu_0}^2,{\mu_1}^2,{\mu_2}^2)\left[{\cal E}_{0}^2-{\widetilde{C}_{\psi} \over  \widetilde{\Omega}_{\psi}^2 + \widetilde{C}_{\psi} \widetilde{A}}{\mu_0}^2\right]}\right\}
\end{equation}

Note the role of the ``fake square''
\begin{equation}
{\cal F}({\mu_0}^2,{\mu_1}^2,{\mu_2}^2) = {\mu_0}^4+{\mu_1}^4+{\mu_2}^4- 2{\mu_0}^2{\mu_1}^2-2{\mu_1}^2{\mu_2}^2-2{\mu_2}^2{\mu_0}^2 
\end{equation}
that is ubiquitous in 3-body phase space. Note that ${\cal F}({\mu_0}^2,{\mu_1}^2,{\mu_2}^2)\ge 0$ for $\mu_0\ge \mu_1+\mu_2$ as required by standard kinematics considerations. In the symmetric case $\mu_1=\mu_2=\mu\le\mu_0/2$ one finds ${\cal F}({\mu_0}^2,{\mu}^2,{\mu}^2)= \mu_0^4 - 4 \mu_0^2\mu^2 = \mu_0^2(\mu_0^2 - 4\mu^2)\ge 0$.

Note also that $\widetilde{A}>0$ and 
$\widetilde{\Omega}_{\psi}>0$ while $\widetilde{C}_{\psi} >0$ for $r>r_e$ and $\widetilde{C}_{\psi} <0$ for $r<r_e$.



The efficiency $\eta$ of Penrose process for JMaRT is given by the energy ${\cal E}_1- {\cal E}_0$ gained by the `probe' particle escaping to infinity with respect to energy of the incoming particle ${\cal E}_0$. As a function of the radial decay point, that we have identified with the radial turning point $r^*$, implicitly determined by the choice of ${\cal E}_0$ and ${\cal J}_0$, {\it viz.}  
\begin{equation}
\eta(r^*) = {{\cal E}_1-{\cal E}_0\over {\cal E}_0} = - {{\cal E}_2\over {\cal E}_0} =-\frac{1}{2 {\mu_0}^2}\left\{({\mu_0}^2-{\mu_1}^2+{\mu_2}^2)
+\sqrt{{\cal F}({\mu_0}^2,{\mu_1}^2,{\mu_2}^2)\left[1-{\widetilde{C}_{\psi} \over  \widetilde{\Omega}_{\psi}^2 + \widetilde{C}_{\psi} \widetilde{A}}{{\mu_0}^2\over{\cal E}_0^2}\right]}\right\}
\end{equation}

\begin{figure}[t!]
    \centering
    \includegraphics[width=\textwidth]{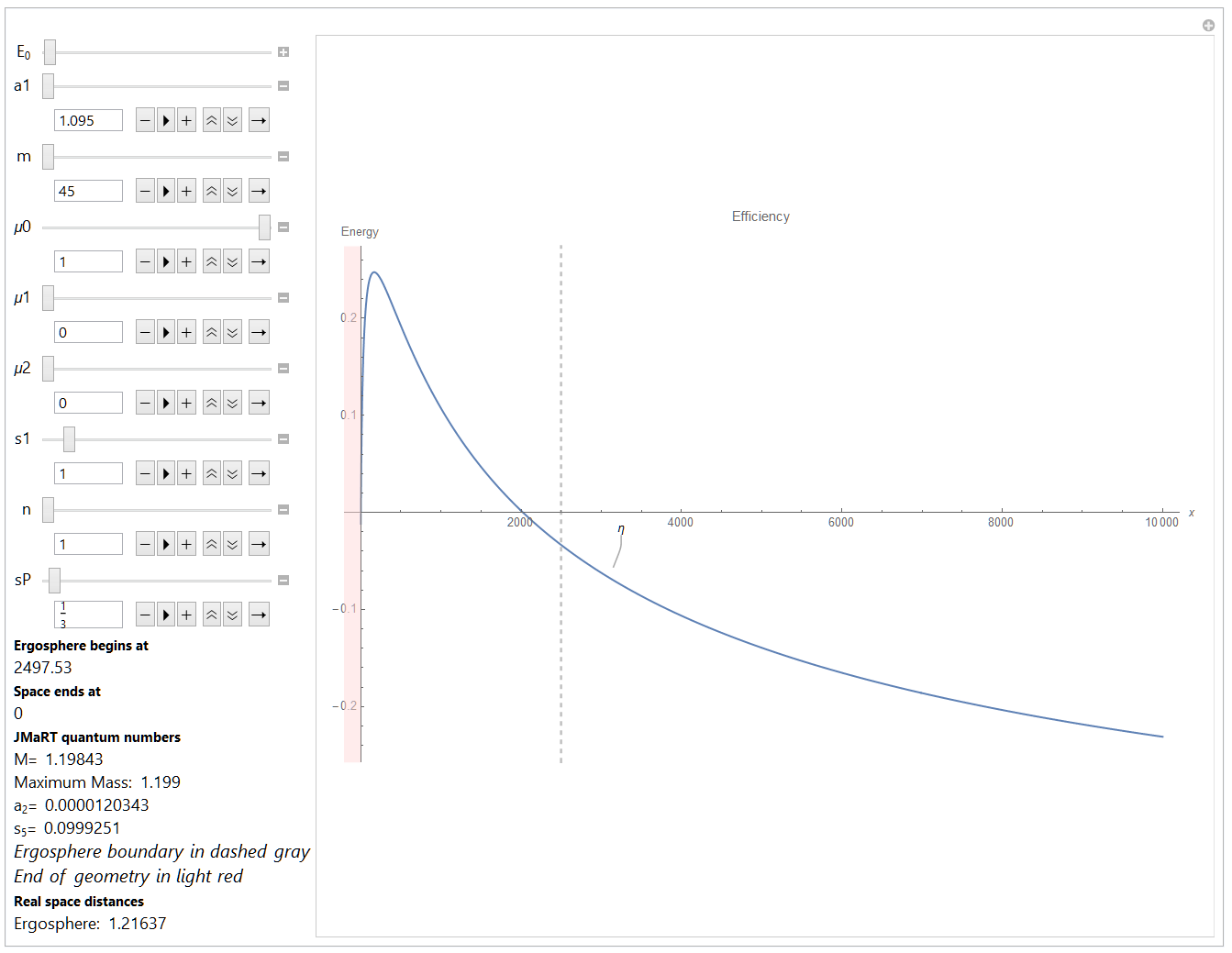}
    \caption{Efficiency for a choice of quantum numbers as in Figs.\ref{fig:NER1}, \ref{fig:NER2}.}
    \label{fig:ETA1}
\end{figure}

The efficiency is negative when $r^*>r_e$ as evident from the plot in Fig. \ref{fig:ETA1}.

For some choice of the parameters, $\eta$ is larger than one (Figs. \ref{fig:ETA2}, \ref{fig:ETA3}). In general, contrary to what happens for rotating BHs, reviewed in the Appendix, there is no upper bound on $\eta$. This looks particularly promising for the acceleration of UHECR, including strangelets, by non-BPS rotating fuzz-balls that should replace putative rotating BHs of the kind found by Kerr. 

The case of JMaRT should be taken as a toy model in many respects. First of all the relevant dynamics is at least five-dimensional. Second, though non-BPS, the charges play a crucial role in the very existence of the solution that should be thought of as some excited state of a BPS configuration. 
Last but not least, achieving phenomenologically reasonable values for the mass and angular momenta require extrapolation to very large charges that may look rather unnatural.

\begin{figure}
    \centering
    \includegraphics[width=\textwidth]{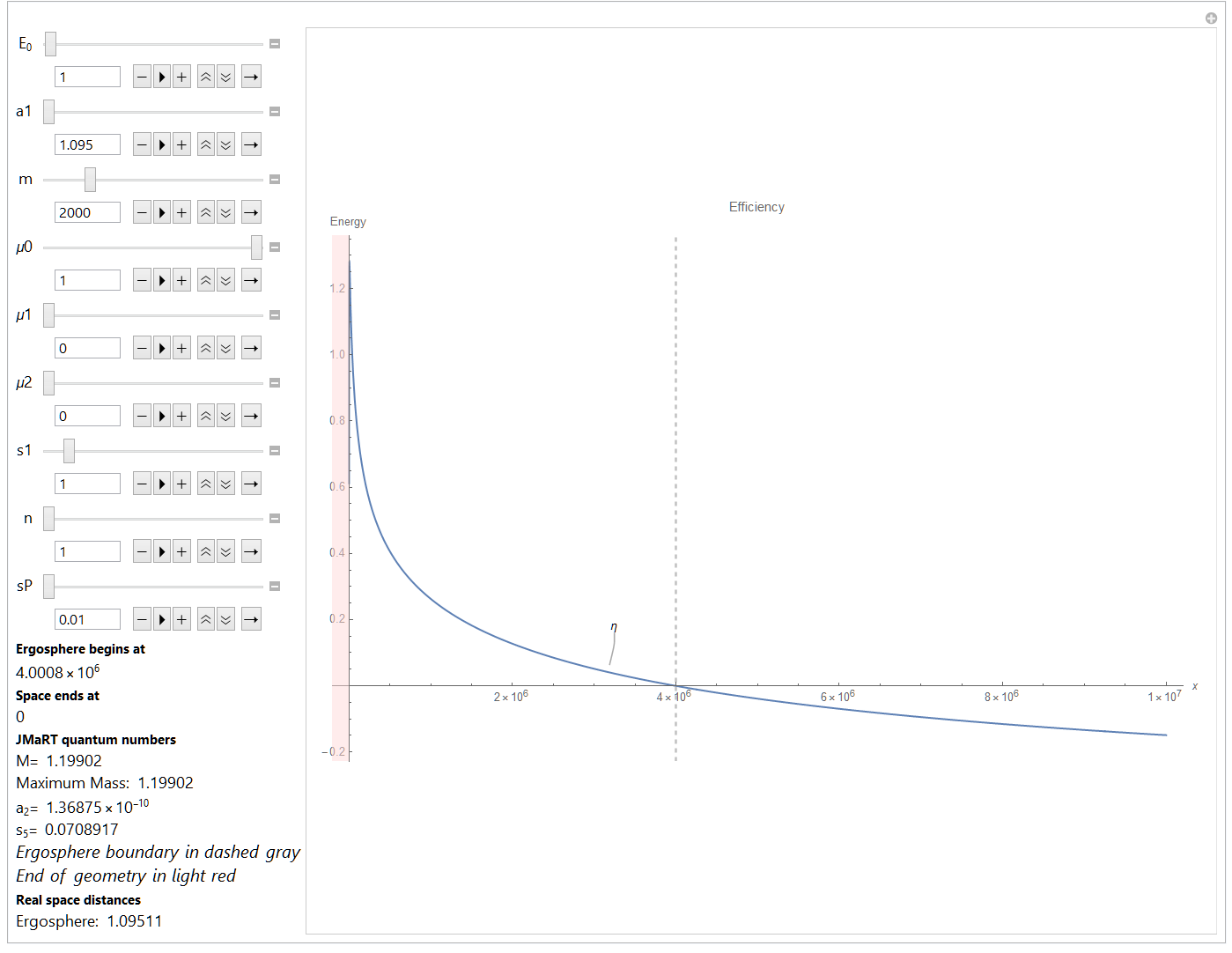}
    \caption{Efficiency for another choice of quantum numbers, shown on the right. It peaks at $\eta_{max}\simeq 1.3$.}
    \label{fig:ETA2}
\end{figure}

\begin{figure}
    \centering
    \includegraphics[width=\textwidth]{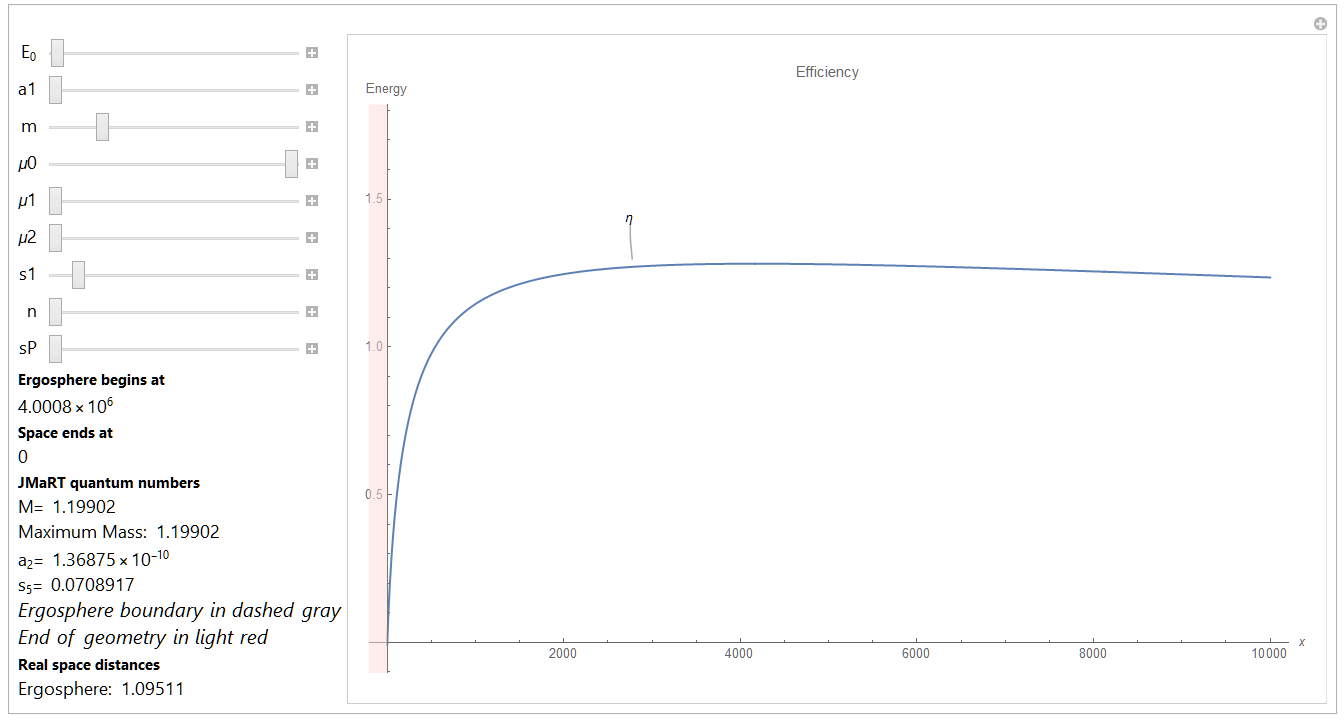}
    \caption{A close-up of the peak in $\eta_{max}$ as claimed in Fig.\ref{fig:ETA2}.}
    \label{fig:ETA3}
\end{figure}

\section{Conclusions and outlook}

After summarising the results of our present analysis, we would like to draw our conclusions and identify directions for future investigation on the subject.

We have shown that the Penrose process can take place not only in singular rotating (Kerr) BHs but  also in smooth horizonless geometries that are expected to represent the micro-states of (charged) rotating BHs.  The common and crucial feature being the presence of an ergoregion. We have considered the case of JMaRT, which is in a loose sense a non-BPS fuzzball in $D=5$. Actually it is over-rotating wrt to {\it classical} BHs with the same mass and charges so it is only a gravitational soliton. Anyway, we took it as a toy model for our analysis and computed the efficiency $\eta$ of a {\it non-collisional} Penrose process in rotating geometries of this kind. Contrary to the case of Kerr BHs, reviewed in the Appendix, $\eta$  is not bounded from above and depends in a highly non-trivial fashion of the parameters of the fuzz-ball as well as on the masses of the probe and of the fragments and above all on the `radial' position where the decay occurs. As expected the efficiency is positive only if the in-falling particle is counter-rotating and the splitting happens inside the ergo-region. 

In order to make quantitative predictions on the relevance of such a mechanism for the acceleration of UHECR and in particular strangelets one should estimate the distribution of such or similar objects in our galaxy / universe as well as of rotating fuzz-balls with large enough mass and angular momentum to be useful as  cosmic slings thus allowing one to reach the GZK cutoff energy of the UHECRs. Sling-shot by other small magnetized objects such as white dwarves, neutron stars (and quark stars) has been proposed by Blandford \& Znajek, Berti, Brito \& Cardoso, Banados and West. We plan to address these and related issues in the near future \cite{MBMCGRinprepmeuso}.

For the time being, we would like to comment on upper limits that MINI-EUSO can set on the flux of strangelets and on the Penrose mechanism for their acceleration derived in Section 3 for non-BPS fuzz-balls and reviewed in the Appendix for Kerr BH. 

MINI-EUSO   is  an instrument to be placed inside the International Space Station (ISS), looking toward  the Earth from a nadir-facing window in the Russian Zvezda module \cite{minieuso}. The main telescope employs a Fresnel optics with a Multi-Anode-photomultiplier (MAPMT) focal surface ($48\times 48$ pixels), with ancillary cameras in the Near-Infrared and Visible regions. Mini-EUSO will map the earth in the UV range (300-400 nm) with a spatial resolution of 6.11 km and a temporal resolution of 2.5 $\mu s$, searching  for Ultra-High Energy Cosmic rays $E>5\cdot 10^{20}\ \text{eV}$ and studying  a variety of atmospheric events such as transient luminous events (TLEs), bioluminescence and  meteors.
The spatial and temporal sampling of the detector allows searching for strange quark matter tracks in the atmosphere, discriminating them from meteors from the light curve (intensity and speed). An estimation of the upper limit of flux which can be posed by one month of night observations of  MINI-EUSO is about $10^{-21} cm^{-2}s^{-1}sr^{-1}$ in the mass range above $5\cdot 10^{24}\ \text{GeV}/c^2$ \cite{eusometeors}.  

\vspace{1cm} 

{\large \bf Acknowledgments}

\vspace{4mm}

We would like to thank Guillaume Bossard, Dario Consoli, Pietro Fr\'e, Giorgio Di Russo, Alfredo Grillo, Maurizio Firrotta for useful discussions and comments.
M.~B. also acknowledges CERN for hospitality during completion of this work. 
M.~B. was partially supported by the MIUR-PRIN contract 2015MP2CX4002 {\it ``Non-perturbative aspects of gauge theories and strings''}.

\section*{Appendix: Penrose process for rotating Kerr BHs}



For comparison with the more laborious case of non-BPS fuzz-balls, represented by JMaRT solutions, let us review how Penrose process can take place in the Kerr metric.

The Kerr black hole is axially symmetric and is characterized
by two parameters: mass $M$ and angular momentum $J = M a$, with $a\le M$. Setting $G_N=1$, the line element in Boyer-Lindquist coordinates reads
\begin{equation} 
\mathrm { d } s ^ { 2 } = - \frac { \Delta - a ^ { 2 } \sin ^ { 2 } \theta } { \rho ^ { 2 } } \mathrm { d } t ^ { 2 } - 2 a \frac { 2 M r \sin ^ { 2 } \theta } { \rho ^ { 2 } } \mathrm { d } t \mathrm { d } \phi  
+ \frac { \left( r ^ { 2 } + a ^ { 2 } \right) ^ { 2 } - a ^ { 2 } \Delta \sin ^ { 2 } \theta } { \rho ^ { 2 } } \sin ^ { 2 } \theta \mathrm { d } \phi ^ { 2 } + \frac { \rho ^ { 2 } } { \Delta } \mathrm { d } r ^ { 2 } + \rho ^ { 2 } \mathrm { d } \theta ^ { 2 } 
\end{equation}\label{eq:KerrMetric}
where  $x=\sqrt{r^2 +a^2}\sin\theta\cos\phi$, $y=\sqrt{r^2 +a^2}\sin\theta\sin\phi$, $z=r\cos\theta$ and 
\begin{equation} 
\Delta   = r ^ { 2 } - 2 M r + a ^ { 2 } \qquad , \qquad  
\rho ^ { 2 } = r ^ { 2 } + a ^ { 2 } \cos ^ { 2 } \theta 
\end{equation}
In this coordinate system, surfaces with constant $t$ and $r$ are deformed two-spheres. The metric for $a = 0$ coincides with Schwarzchild metric. In contrast to the latter, however, there is an off-diagonal term \begin{equation}
g _ { t \phi } = - a \frac { 2 M r \sin ^ { 2 } \theta } { \rho ^ { 2 } }
\end{equation}
that is responsible for the `gravitational' dragging of inertial frames caused by the rotation of the source. 
In practice a particle dropped `straight' in from infinity, i.e. with ${\cal J} \equiv P_\phi =0$ is `dragged' just by the influence of gravity so that it acquires an angular velocity $\omega$ in the same sense as that of the source. For the Kerr metric, $\omega$ has the same sign as $a=J/M$. This effect weakens with the distance as $1 / r ^ { 3 } $.

Kerr metric presents a singularity, a horizon and an ergo-sphere. 

The singularity is a ring located in the equatorial plane $\theta=\pi/2$, at $r=0$ {\it i.e.} $z=0$ and $x^2+y^2=a^2$.

The singularity is cloaked by a horizon where $g _ { r r } = \infty$,  {\it i.e.} $\Delta = 0$ that corresponds to the radius
\begin{equation}
r _ { + } = M + \sqrt {  M ^ { 2 } - a ^ { 2 } }
\end{equation}\label{eq:KerrHorizon}
The `ergo-sphere' can be identified as the surface where the norm of the time-like Killing vector $V_t = \partial_t$ vanishes. It is also called the `static limit', since inside it no particle can remain at fixed $r , \theta , \phi$. From (\ref{eq:KerrMetric}) one finds 
\begin{equation} 
||V_t||^2 = g_{\mu\nu} V^\mu_t V^\nu_t = g_{tt} = - \frac { \Delta - a ^ { 2 } \sin ^ { 2 } \theta } { \rho ^ { 2 } } = 0 \quad {\rm for} \quad \Delta = a ^ { 2 } \sin ^ { 2 } \theta
\end{equation}
that means\footnote{For later use, note that $r _ {e}=2M$ for $\theta=\pi/2$.}
\begin{equation}
    r _ {e}(\theta) = M + \sqrt {  M ^ { 2 } - a ^ { 2 } \cos ^ { 2 } \theta }
\end{equation}
The ergo-sphere lies outside the horizon except at the poles, $\theta=0,\pi$, where 
they touch each other. In the ergo-region, inside the ergo-sphere, all particles, including photons, must rotate with the hole since $g^{tt} > 0$. The presence of the ergo-region allows Penrose process to take place as we will see momentarily.



Focussing for simplicity on geodesics in the equatorial plane $\theta =\pi/2$ allows to write the restricted metric in the compact form 
\begin{equation}
    {{d}}s^2 = -A\ {{d}} t ^ { 2 } + C \ {{d}} \phi ^ { 2 } + 2\Omega\ {{d}} t {{d}} \phi + B\ {{d}} r ^ { 2 }  
\end{equation}\label{eq:metricABC}
where 
\begin{equation}
A = 1- {2M\over r} \quad , \quad 
B= \left( 1- {2M\over r}  + {a^2\over r^2} \right)^{-1} \quad ,\quad C= r^2 + a^2 + {2M\over r} a^2 \quad , \quad \Omega= -  {2Ma\over r} 
\end{equation}
Computing the conjugate momenta $P_\mu = g_{\mu\nu} \dot{x}^\nu$ and setting\footnote{We denote the angular momentum of the probe by ${\cal J}$ in order to avoid confusion with the angular momentum of the Kerr BH, denoted by $J=Ma$.}
\begin{equation}
P _ { t } = - {\cal E} \quad , \quad P _ { \phi } = {\cal J} \quad , \quad P _ { r } = {\cal P}
\end{equation}
one finds
\begin{equation}
\mathcal{H}=\frac { 1 } { 2 B } {\cal P} ^ { 2 } + \frac { 1 } { 2 }\ \frac { 1 } { A C + \Omega ^ { 2 } } \left[ - C {\cal E}^ { 2 } - 2 \Omega {\cal E}{\cal J} + A {\cal J} ^ { 2 } \right] = -{\mu^2\over 2}
\end{equation}
where $\mu$ is the mass of the probe particle. The geodesic is null for $\mu=0$.
Resolving for the radial momentum ${\cal P}$ in terms of ${\cal E}$ and ${\cal J}$ one finds
\begin{equation}
{\cal P}^{ 2 } = \frac { B } { A C + \Omega ^ { 2 } } \left[ C {\cal E}^ { 2 } + 2 \Omega {\cal E}{\cal J} - A {\cal J}^ { 2 } + {{\mu^2}}  \left( A C + \Omega ^ { 2 } \right) \right] = 
\frac { B C } { A C + \Omega ^ { 2 } }({\cal E} - {\cal E}_+)({\cal E} - {\cal E}_-) \ge 0
\end{equation}
where the `effective potentials' read
\begin{equation}
{\cal E}_{\pm} = {- \Omega{\cal J} \pm \sqrt{ (\Omega^2 + AC)({\cal J}^2-C\mu^2)} \over C} \end{equation} 
Since ${ B C }/A C + \Omega ^ { 2 }\ge 0$ outside the horizon either 
${\cal E} \ge {\cal E}_{+}> {\cal E}_{-}$ or ${\cal E} \le {\cal E}_{-}< {\cal E}_{+}$. ${\cal E}_{\pm}$ determine allowed negative-energy regions. 
For co-rotating particles (${\cal J}a\ge 0$) ${\cal E}^{\uparrow\uparrow}_{+}$ is always positive, while while
for counter-rotating particles (${\cal J}a\le 0$) ${\cal E}^{\uparrow\downarrow}_{+}$ becomes negative  inside the ergo-sphere ($r_e=2M$).

As mentioned above, if a positive energy counter-rotating particle enters the ergo-sphere it acquires negative energy and `decays' into two or more products, one of which has negative energy and falls into the horizon, then the particle that escapes may have more energy than the initial particle. In this way Kerr BH loses mass angular momentum.

Following \cite{Chandra, Shutz}, we now review the efficiency of the Penrose process in Kerr BH.
 
For simplicity we will assume that the in-falling massive particle has ${\cal E}=\mu$ (`rest mass'), that the products are massless scalars (no spin) $\mu_1=\mu_2=0$ and that the decay takes place at the turning point $r=r^*$ (with $r_H<r^*<r_e=2M$) where $P_r=0$.  
Since energy and angular momentum are conserved, we have
\begin{equation}
{\cal E}_1+{\cal E}_2={\cal E} \qquad , \qquad {\cal J}_1+{\cal J}_2={\cal J}
\end{equation}
For massless particles ${\cal J}_i = \alpha{\cal E}_i$, while 
${\cal J}= \beta{\cal E}$ for the massive one with $\alpha$ and $\beta$ depending on the turning point $r^*$ where the decay/splitting takes place. The second equation then simplifies drastically to
\begin{equation}
\alpha_1 {\cal E}_1+ \alpha_2 {\cal E}_2=\beta {\cal E}
\end{equation}
Solving the `linear' system one has
\begin{equation}
{\cal E}_1= {\beta -\alpha_2 \over \alpha_1 -\alpha_2} {\cal E}
\qquad , \qquad {\cal E}_2= {\beta -\alpha_1 \over \alpha_2 -\alpha_1} {\cal E}
\end{equation}
Taking the positive branch for ${\cal E}={\cal E}_{+}$ and ${\cal E}_1={\cal E}_{1,+}$ and the negative branch for ${\cal E}_2={\cal E}_{2,-}<0$, one gets ${\cal E}_1>{\cal E}$
The efficiency of the process can be estimated in the following way. Since
$\Delta{\cal E}={\cal E}_1-{\cal E} = -{\cal E}_2$ is the gained energy, the efficiency of Penrose process as a function of $r^*$ is given by  
\begin{equation}
\eta(r^*)=\frac{{\cal E}_1-{\cal E}}{{\cal E}} = \frac{-{\cal E}_2}{{\cal E}} ={\beta -\alpha_1 \over \alpha_1 -\alpha_2} = {1\over 2} \left(\sqrt{2M\over r^*}- 1 \right) \le {1\over 2}(\sqrt{2}- 1 )
\end{equation}
since $r_H\le r^*\le 2M=r_e(\theta=\pi/2)$ for the very process to take place. In fact the maximum is reached when $r^*=r_H = M +\sqrt{M^2-a^2}$ and $a=M$ (extremal Kerr BH).

\end{document}